\documentclass{ceurart}

\usepackage{listings}
\usepackage{enumitem,kantlipsum}
\usepackage{adjustbox}
\usepackage{bibunits}

\begin{document}

\copyrightyear{2026}
\copyrightclause{Copyright for this paper by its authors.
  Use permitted under Creative Commons License Attribution 4.0
  International (CC BY 4.0).}

\conference{ISWC 2026 Companion Volume, October 25-29, 2026, Bari, Italy}

\title{Ontology Interoperability: A Comprehensive Framework for Industry-Scale Applications}

\author[1]{Zhangcheng Qiang}[
orcid=0000-0001-5977-6506,
email=qzc438@gmail.com,
]
\cormark[1]

\cortext[1]{Corresponding author.}

\begin{abstract}
Different ontologies with conflicting and overlapping concepts cause havoc in the design, develop, and deploy phases of ontology-driven applications. In this work, we propose a comprehensive ontology interoperability framework for industrial-scale ontology-driven applications. The framework employs three state-of-the-art semantic techniques in different phases of the ontology engineering (OE) life cycle: ontology design patterns (ODPs) in the design phase, ontology matching and versioning (OM\&OV) in the develop phase, and ontology validation (OVA) in the deploy phase, to achieve better ontology interoperability.
\end{abstract}


\maketitle

\section{Problem Statement}

An ontology is ``a formal and explicit specification of a shared conceptualisation''~\citep{gruber1993translational}. They are commonly used to define terms and relations in a domain of interest, for example, web pages, catalogues, map search, electronic health records, medical billing, and biological taxonomy. Ontologies improve semantic interoperability for data integration and knowledge sharing. But is the ontology itself interoperable?

Ontologies are crafted by domain experts based on their understanding of the domain of application. There is no guarantee that ontologies in the same domain share the same conceptualisation. Describing a person can use their basic profile (e.g. name, age, and gender), social relationships (e.g. parent-child and friendship), or professional roles (e.g. student and employee). The terminology used in the ontologies may vary.  For example, one ontology can use ``first name'' and ``last name'' to describe a person's full name, and another ontology can use ``given name'' and ``family name'' instead. These concepts need to be harmonised in order for data conforming with each ontology to be shared in applications. Therefore, ontology interoperability has become a critical issue that limits sharing data and the reuse of ontologies.

\section{Importance}

Knowledge graphs (KGs) can use ontologies as the backbone. In particular, ontologies serve to provide extensive information for interpreting KGs. KGs are commonly formed as a set of triples, each consisting of a subject, a predicate, and an object. KGs are sometimes not easy to understand. For example, the KG triple (Sydney, travelsTo, Sydney) may cause confusion. We can use an ontology to help. With an ontology that defines a class expression (Person, travelsTo, Place), one can infer that the subject ``Sydney'' refers to a person who is travelling, while the object ``Sydney'' refers to the place being travelled. The KG triple (Sydney, travelsTo, Sydney) describes a filming event in which the American actress Sydney Sweeney travels to the city of Sydney~\citep{sweeney2023}.

For large-scale KGs, the use of ontologies has become the general practice for enabling semantic interoperability among KGs. With the increasing size and number of ontologies, the lack of interoperability among ontologies has become apparent and hampers their use in many real-world applications. Missing, overlapping, and conflicting concepts in different ontologies impede data exchange and integration. While several approaches have been proposed for ontology interoperability, they often operate on a case-by-case basis. However, ontology interoperability is a complex issue that can arise at any phase of the ontology engineering (OE) cycle. Currently, there is no streamlined approach to improving ontology interoperability throughout the OE life cycle.

\section{Related Work}

There are three popular approaches to achieve ontology interoperability.

\begin{itemize}[wide, noitemsep, topsep=0pt, labelindent=0pt]

\item \textbf{Ontology design patterns (ODPs)} identify common ontology fragments across different ontologies~\citep{gangemi2009ontology}. ODPs use a hub-and-spoke matching approach. Once the ODP is determined, each subsequent ontology will conform to this ODP, and the ODP serves as a high-level abstract model to guide the design of these new ontologies. ODPs have some limitations. Firstly, ODPs are conceptual models and do not prescribe the identifiers for entities. This means that it is not possible to control terminology, and ontology creators remain free to make their own choices. Secondly, ODPs are commonly handcrafted based on existing ontologies. They are static and tend not to capture changes over time. Thirdly, the choice of ODPs is entirely reliant on one's experience and preferences; the favoured ODPs may be overwhelming, overlapping, or not applicable for a given KG.

\item \textbf{Ontology matching and versioning (OM\&OV)} aim to find mappings between different ontologies~\citep{euzenat2007ontology} or different versions of the same ontology~\citep{klein2001ontology}. OM\&OV conform to a pairwise matching approach. This means that each mapping is a pairwise process to find the correspondences between two ontologies. OM\&OV are typically resource-intensive. The complexity of pairwise mapping is proportional to the product of the number of entities in the two ontologies. The pairwise mapping process also needs to be repeated multiple times if multiple ontologies exist. Moreover, OM\&OV tasks focus on conceptual-to-conceptual and instance-to-instance matching, and less attention has been paid to the alignment between conceptual models and data instances.

\item \textbf{Ontology validation (OVA)} aims to answer the question ``Do we build the ontology correctly?'', in other words, to ensure that the ontology is extrinsically meaningful. In this work, we use OVA to validate interoperability among entities defined in different ontologies. OVA can be performed at the conceptual and instance levels. The former often requires domain experts to manually validate the proposed mappings, whereas the latter typically uses real-world data to validate the effectiveness of an ontology. Today, with the increasing number of domains and applications using ontologies, it is not easy to find domain experts and experts may also have knowledge boundaries for OVA tasks. On the other hand, validating the ontology on KGs is challenging. To determine which candidate ontology best fits the given KG, the KG needs to be modelled for each candidate ontology. This process essentially involves aligning the candidate ontologies, informed by the KG under study.

\end{itemize}

\section{Research Question(s) and Hypotheses}

In this work, we aim to build a comprehensive framework for ontology interoperability. We review three state-of-the-art semantic web techniques: ontology design patterns (ODPs), ontology matching and versioning (OM\&OV), and ontology validation (OVA). By analysing the pros and cons of these techniques, we hypothesise that using these three techniques across different phases of the ontology engineering life cycle can foster strengths and circumvent weaknesses, thereby streamlining a workflow model for ontology interoperability.

\begin{figure}[htbp]
\centering
\includegraphics[width=0.7\linewidth]{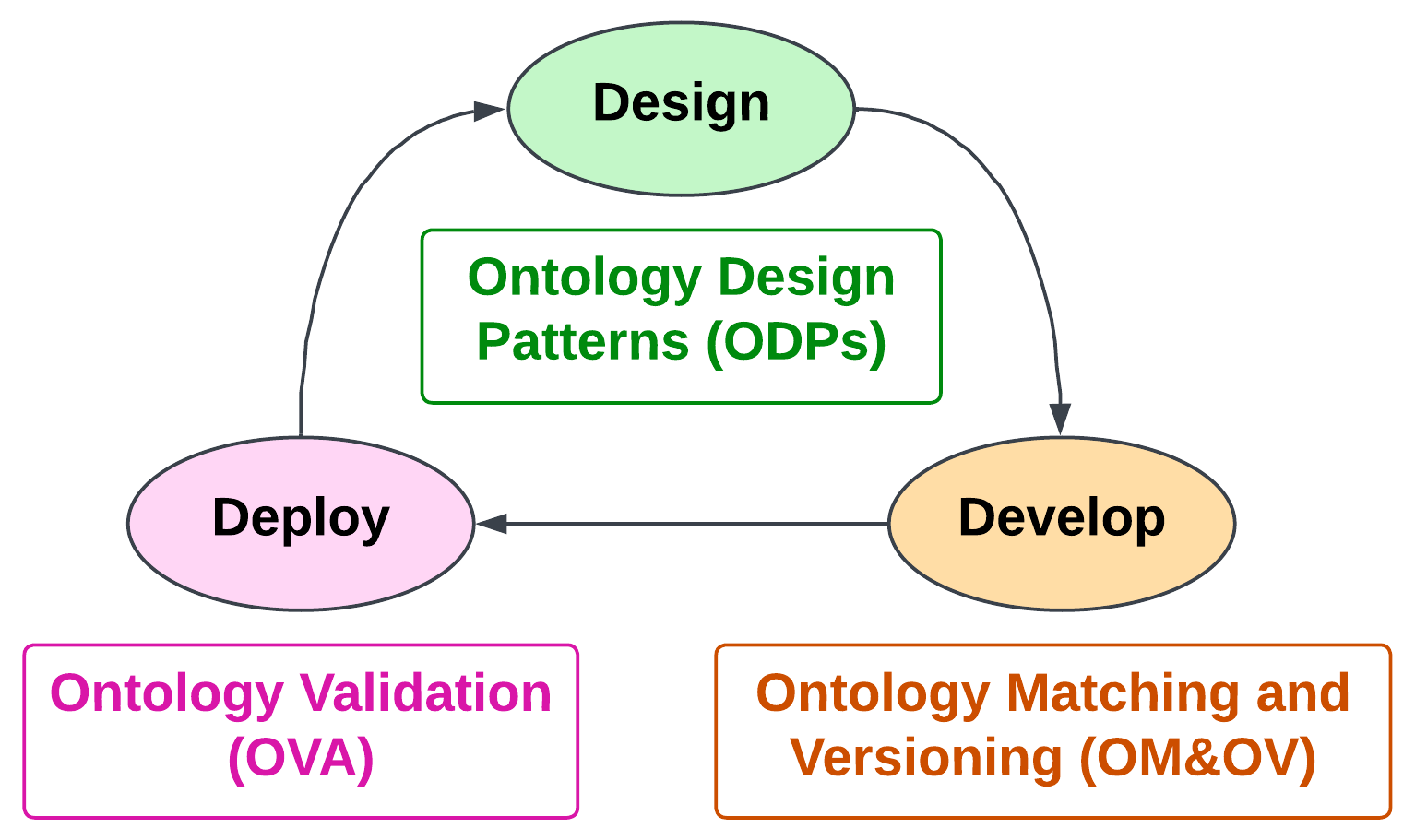}
\caption{An overview of the ecosystem. We propose to use ODPs in the design phase, OM\&OV in the develop phase, and OVA in the deploy phase to achieve better ontology interoperability.}
\label{fig: overview}
\end{figure}

Figure~\ref{fig: overview} illustrates an overview of the framework. Similar to software engineering, ontology engineering (OE) involves three main phases: design, develop, and deploy. We propose to employ ODPs, OM\&OV, and OVA in different phases. (1) The design phase is responsible for understanding the requirements and creating a blueprint or prototype. In OE, these requirements are collected by competency questions. ODPs can serve as a skeleton to create conceptual models based on competency questions. (2) The develop phase includes turning the prototype into a full product and applying a unit test. In OE, this phase is referred to as ontology construction and evaluation. OM\&OV can be used to evaluate the level of interoperability between two ontologies, or between an ontology and its previous version. (3) The deploy phase usually refers to implementing the product in the real world. In OE, we refer to this as validating the ontology against human values and real-world data instances via OVA, thereby enhancing the design of the ontology in the next cycle.

\section{Preliminary Results}

Driven by the emergence of efficient digital transformation for Industry 4.0, numerous building ontologies have been developed over the past two decades. While building concepts are complex and evolve rapidly, ontology interoperability has become a prominent problem in the building domain. Ontologies used in the building domain encompass diverse concepts and levels of abstraction, and building information systems based on these ontologies face communication hurdles~\citep{qiang2023systematic}.

We use a running example to illustrate the application and value of our proposed framework for ontology interoperability. The running example concerns sensor observations in the building, for example, ``A sensor is installed in the room to monitor the room temperature.'' The ontologies used in this case are required to capture several main concepts, including sensor, room, temperature, and sensor observation. In the real world, building such an ontology requires considerable effort. The proposed ontology should capture all concepts without overlap or omission; moreover, it should be compact and easy to migrate to another ontology if the system's vendors change over the years.

\subsection{ODPs in the Design Phase}

\begin{figure}[!t]
\centering
\includegraphics[width=\linewidth]{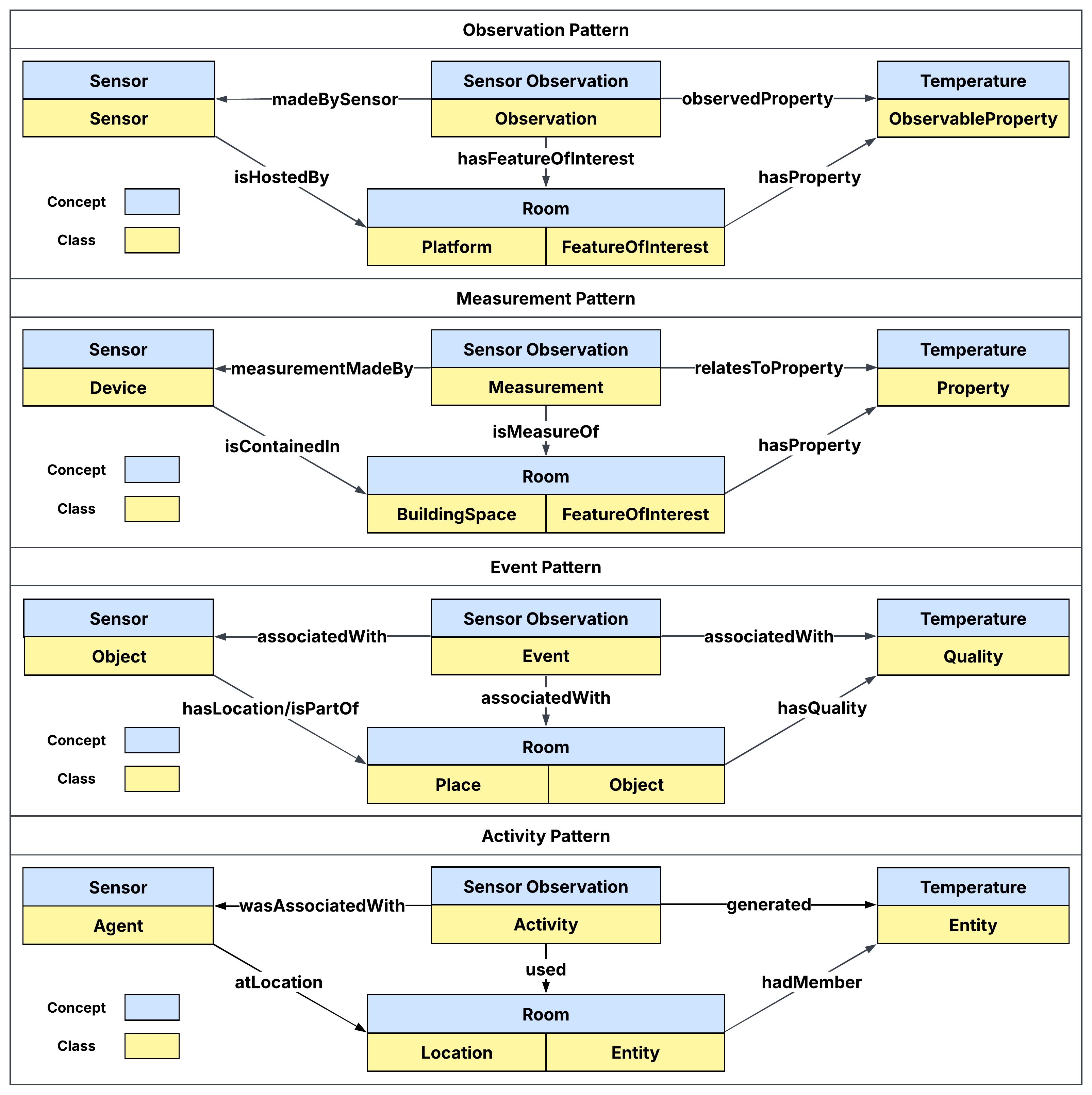}
\caption{Modelling sensor observation using different ODPs.}
\label{fig: observation}
\end{figure}

Ontologies can be built from scratch, but such ontologies are likely to have more customised concepts and limited interoperability. It is commonly recommended to reuse pre-existing ODPs. Figure~\ref{fig: observation} illustrates that four classical ODPs can be used in this case to model the example of sensor observation from different perspectives.

\begin{enumerate}[wide, noitemsep, topsep=0pt, labelindent=0pt, label=(\roman*)]
\item \textit{Observation Pattern} (used in SOSA~\citep{janowicz2019sosa} and SSN~\citep{compton2012ssn}). Sensor observation is defined as a type of observation. The observation is made by the sensor, which has a feature of interest in the room and the observed property temperature.
\item \textit{Measurement Pattern} (used in the early version of SAREF~\citep{daniele2015created}). Sensor observation is defined as a measurement. The measurement is made by the sensor, which is a measure of the room related to the property temperature.
\item \textit{Event Pattern} (used in DUL~\citep{presutti2016dolce+}). Sensor observation is defined as an event. The event is associated with the sensor, room, and temperature.
\item \textit{Activity Pattern} (used in PROV-O~\citep{w3c-prov}). Sensor observation is defined as an activity. The activity is associated with the sensor that uses the room and generates a temperature measurement.
\end{enumerate}

While all four of these ODPs can effectively capture the main concepts, significant differences in perspective need to be considered, as shown below.

\begin{enumerate}[wide, noitemsep, topsep=0pt, labelindent=0pt, label=(\roman*)]
\item The definitions of the concepts can differ. Temperature is defined as a property of the room in both the \textit{Observation Pattern} and the \textit{Measurement Pattern}, but in the \textit{Event Pattern} it is treated as a quality measure of the room. \textit{Activity Pattern} defines the temperature as a member of the room.
\item The levels of expressiveness are different. Unlike \textit{Observation Pattern}, \textit{Measurement Pattern}, and \textit{Activity Pattern}, which can only describe a sensor that is installed in the room, the \textit{Event Pattern} can distinguish between a structural attachment (i.e. isPartOf) and a spatial presence (i.e. isLocatedIn).
\end{enumerate}

\subsection{OM\&OV in the Develop Phase}

\begin{figure}[htbp]
\centering
\includegraphics[width=\linewidth]{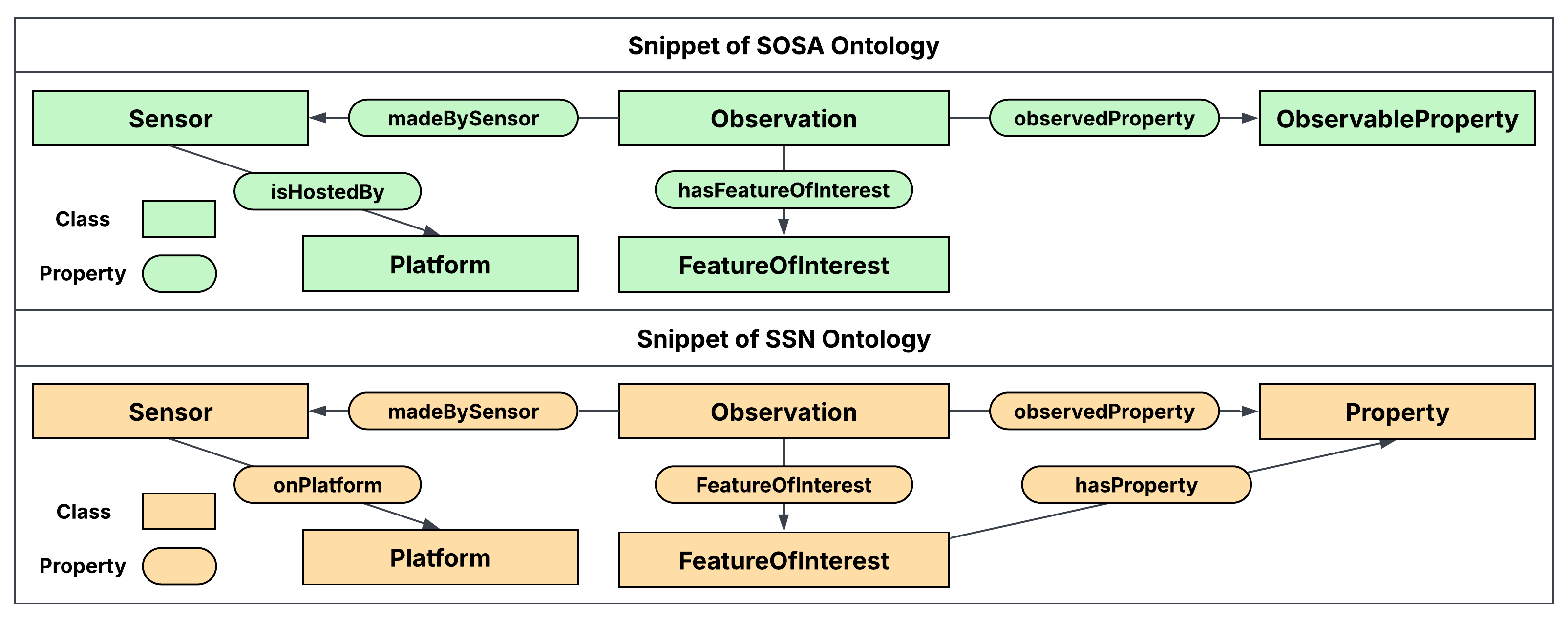}
\caption{Aligning SOSA and SSN.}
\label{fig: ssn&sosa}
\end{figure}

We assume that the \textit{Observation Pattern} is chosen in the design phase and implemented in two ontologies, SOSA and the original SSN~\citep{compton2012ssn,old-ssn} (also called the old SSN in the W3C standard~\citep{w3c-ssn}). As illustrated in Figure~\ref{fig: ssn&sosa}, SOSA and SSN are very similar because they are built from the same ODPs. However, they are not exactly the same. The OM task is to identify aligned concepts across two ontologies. SOSA and an updated SSN were co-developed, with a published alignment between them. Sosa was conceived as a module of the updated SSN (for which there were other modules too). If we consider SOSA to be an updated and reduced version of SSN, then this OM task is also an OV task. We use OM\&OV here because OM and OV can be unified with minor modifications~\citep{qiang2024om4ov}. OM\&OV tasks can be classified into two main categories.

\begin{enumerate}[wide, noitemsep, topsep=0pt, labelindent=0pt, label=(\roman*)]
\item Finding trivial alignments only considers the syntactic matching of the entity name. For example, the mappings between the classes \textit{sosa:Sensor} and \textit{ssn:Sensor}, and the properties \textit{sosa:madeBySensor} and \textit{ssn:madeBySensor} are trivial. We also refer to these mappings as ``seed mappings'' because they serve as the basis for subsequent lexical and semantic matching.
\item Finding non-trivial alignments considers syntactic, lexical, and semantic matching of the entity. For example, the mappings between the classes \textit{sosa:ObservableProperty} and \textit{ssn:Property}, and the properties \textit{sosa:isHostedBy} and \textit{ssn:onPlatform} should be discovered. The alignment should also exclude non-existent and counter-intuitive mappings. For example, \textit{ssn:hasProperty} does not have a direct equivalent mapping in SOSA.
\end{enumerate}

\subsection{OVA in the Deploy Phase}

While the mappings developed by the OM systems are not 100\% reliable, additional human validation is required. A number of experts will sit together to determine whether these mappings are correct. This procedure can be conducted in surveys, interviews, and workshops.

\begin{figure}[htbp]
\centering
\includegraphics[width=\linewidth]{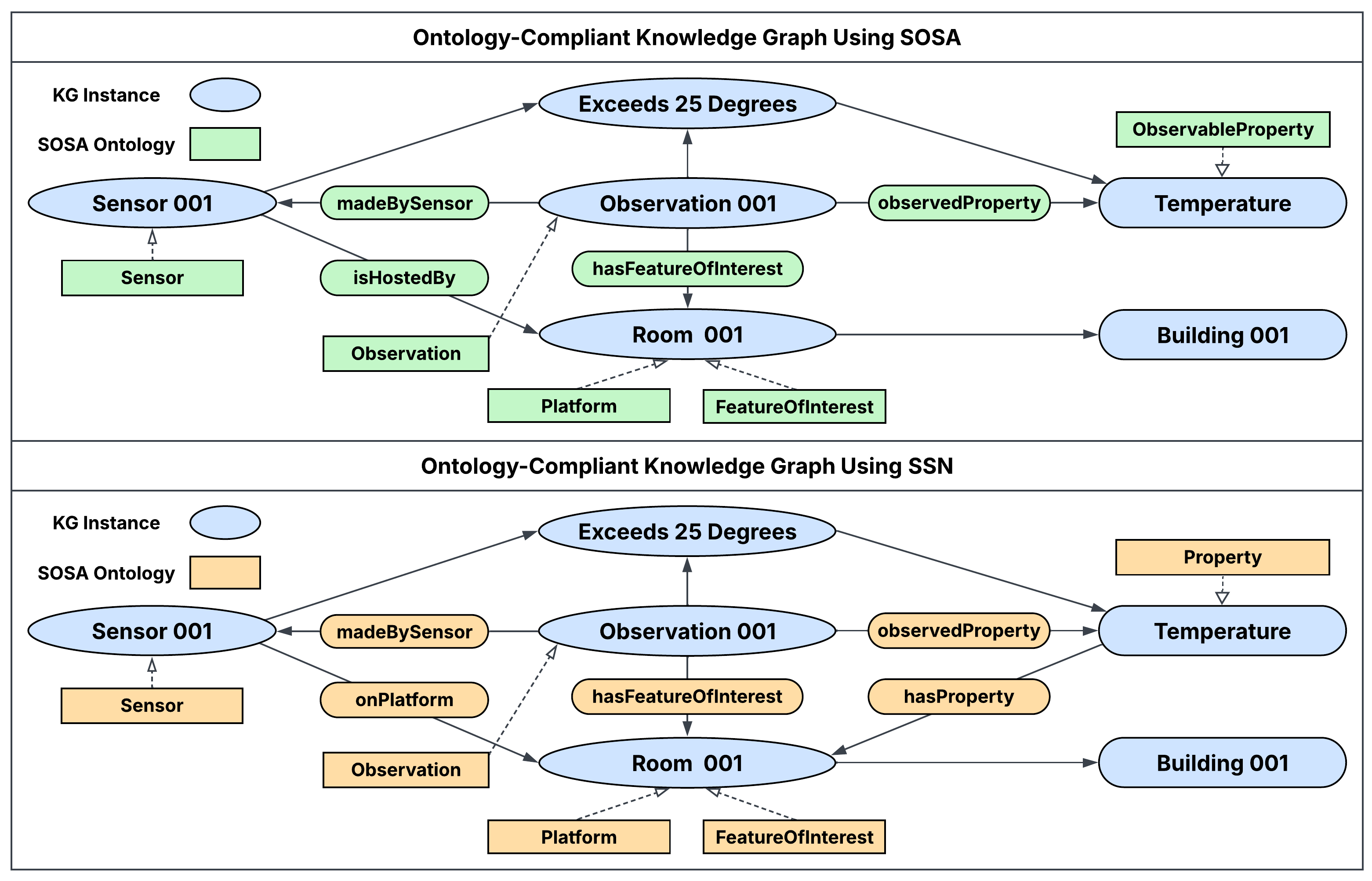}
\caption{Build OCKGs using SOSA and SSN.}
\label{fig: ockg-sosa-ssn}
\end{figure}

Say we have developed SOSA and SSN, along with their reference alignment validated. We now face the question of which ontology to use in real-world KGs. Some concepts used in a KG may not have the corresponding classes and properties in the ontology. Therefore, we apply these two ontologies separately to the KGs and construct ontology-compliant knowledge graphs (OCKGs) for evaluation. The introduction of OCKGs is a novel DOVA approach that uses a pay-as-you-go model to find the most cost-effective ontology fragment that matches the given KGs~\citep{qiang2023ontology}. Figure~\ref{fig: ockg-sosa-ssn} shows the OCKGs based on SOSA and SSN. We can see that both ontologies can capture the general concepts of this sensor observation; however, they fail to capture two specific concepts in the given KG.

\begin{enumerate}[wide, noitemsep, topsep=0pt, labelindent=0pt, label=(\roman*)]
\item The stimulus of the sensor ``Exceeds 25 Degrees''. The sensor only makes an observation when the temperature exceeds 25 degrees.
\item The actual feature of interest ``Building 001''. The sensor observed the temperature of Room 001. It is not a feature of interest, but a sample of the temperature of Building 001.
\end{enumerate}

To model these two new concepts and adequately integrate them with the other parts of a KG, we need to extend the ODP with the \textit{Stimulus-Sensor-Observation Pattern} from SSN and the \textit{Sampling Pattern} from SOSA, as well as make several design choices between aligned mappings (retrieved from~\citep{w3c-ssn,haller2018modular,taylor2019semantic}).

\begin{enumerate}[wide, noitemsep, topsep=0pt, labelindent=0pt, label=(\roman*)]
\item We use \textit{ssn:Property} to replace \textit{sosa:ObservableProperty} because this class is more general and can be applied in a broader context for sensing, including actuation and sampling.
\item We use \textit{sosa:isHostedby} to replace \textit{ssn:onPlatform} because this property is more closely related to describing a sensor physically installed in the room.
\item We use \textit{sosa:hasFeatureOfInterest} to replace \textit{ssn:FeatureOfInterest}. Although there is no restriction for \textit{ssn:FeatureOfInterest} to be used both as a class and as a property, we add the verb ``has'' to distinguish the property from the class for human readability.
\end{enumerate}

\begin{figure}[htbp]
\centering
\includegraphics[width=\linewidth]{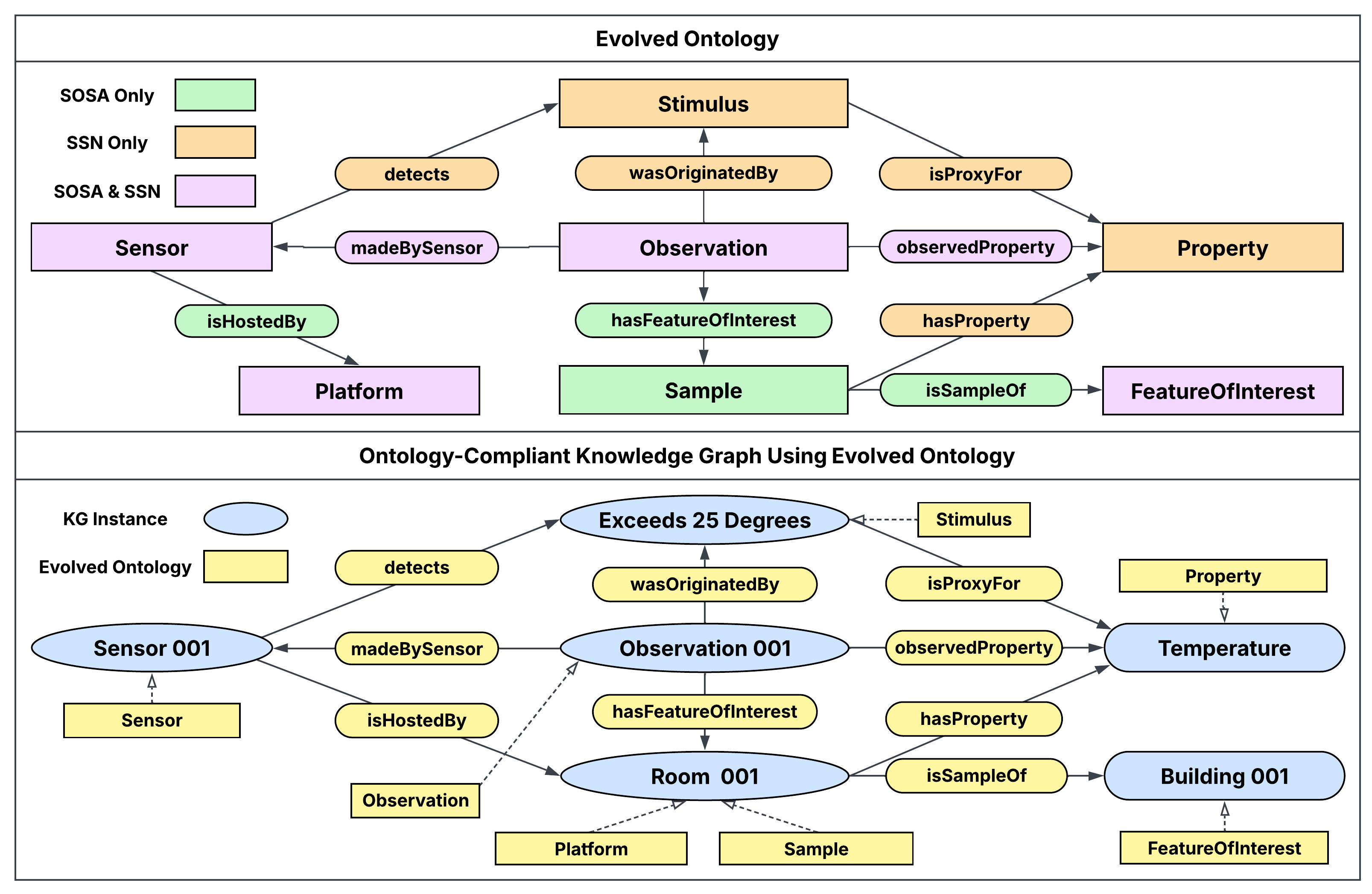}
\caption{Updated ontology and new OCKG.}
\label{fig: odp-new}
\end{figure}

Figure~\ref{fig: odp-new} shows the updated ontology and the OCKG built on the updated ontology. It appears that the updated ontology is better suited to the provided KG instance. In industrial scenarios, an ontology can be deployed in multiple KGs, and the procedure described above can be repeated multiple times. It is also possible to choose other alternative ODPs. For example, the class ``BuildingSpace'' in the \textit{Measurement Pattern} is more suitable to describe a room, and the property ``hasLocation/isPartOf'' in the \textit{Event Pattern} is more appropriate to describe the link between the sensor and the room.

\section{Evaluation}

Well-defined ODPs can serve as reusable ontology fragments, thereby reducing the complexity of OM\&OV tasks. As a complement to ODPs, OM\&OV can align detailed classes, properties, and individuals that are not captured by ODPs. On the other hand, while OCKGs are useful tools for validating ontology interoperability and evolving ODPs, their construction relies heavily on precise mappings produced by OM\&OV systems with additional human-in-the-loop validation. OM\&OV is a critical middleware between ODPs and OVA.

\section{Reflection and Future Work}

Ontology interoperability has been a long-standing issue studied in the Semantic Web community. Since the establishment of the International Semantic Web Conference (ISWC), workshops on ontology design and patterns (WOP) and ontology matching (OM@ISWC) have been the two longest-running workshop proceedings. Ontology validation, which focuses on validating the correctness of the ontology, has been used in many ontology-driven applications.

While there is no preference for one technique over another, these three techniques play different roles in a comprehensive framework for ontology interoperability. We believe that employing ODPs in the design phase, utilising OM\&OV in the develop phase, and using OVA in the deploy phase could significantly improve ontology interoperability throughout the ontology engineering life cycle.

\begin{acknowledgments}
The author is supervised by Kerry Taylor (Australian National University, ANU), Weiqing Wang (Monash University), Jing Jiang (ANU), Subbu Sethuvenkatraman (Commonwealth Scientific and Industrial Research Organisation, CSIRO), Qing Wang (ANU), and Alex Potanin (ANU). The author thanks CSIRO for supporting this project.
\end{acknowledgments}

\section*{Declaration on Generative AI}

During the preparation of this work, the authors used Grammarly in order to grammar and spell check, and to improve the text readability. After using the tool, the authors reviewed and edited the content and take full responsibility for the publication's content.

\clearpage
\bibliography{qiang-bibliography-dc}

\begin{thebibliography}{10}

\bibitem{dul}
{DUL}, n.d.

\bibitem{san}
{SAN}, n.d.

\bibitem{seas}
{SEAS}, n.d.

\bibitem{sosa}
{SOSA}, n.d.

\bibitem{ssn}
{SSN}, n.d.

\bibitem{aranguren2008ontology}
Mikel~Ega{\~{n}}a Aranguren, Erick Antezana, Martin Kuiper, and Robert Stevens.
\newblock Ontology design patterns for bio-ontologies: a case study on the cell cycle ontology.
\newblock {\em {BMC} Bioinformatics}, 9({S-5}), 2008.

\bibitem{giglou2024llms4om}
Hamed Babaei~Giglou, Jennifer D'Souza, Felix Engel, and S{\"o}ren Auer.
\newblock {LLMs4OM}: Matching ontologies with large language models.
\newblock In {\em The Semantic Web: {ESWC} 2024 Satellite Events}, pages 25--35, Hersonissos, Crete, Greece, 2024. Springer.

\bibitem{balaji2016brick}
Bharathan Balaji, Arka Bhattacharya, Gabriel Fierro, Jingkun Gao, Joshua Gluck, Dezhi Hong, Aslak Johansen, Jason Koh, Joern Ploennigs, Yuvraj Agarwal, Mario Berges, David Culler, Rajesh Gupta, Mikkel~Baun Kj\ae{}rgaard, Mani Srivastava, and Kamin Whitehouse.
\newblock Brick: Towards a unified metadata schema for buildings.
\newblock In {\em Proceedings of the 3rd ACM International Conference on Systems for Energy-Efficient Built Environments}, pages 41--50, Palo Alto, CA, USA, 2016. ACM.

\bibitem{arrieta2020explainable}
Alejandro {Barredo Arrieta}, Natalia Díaz-Rodríguez, Javier {Del Ser}, Adrien Bennetot, Siham Tabik, Alberto Barbado, Salvador Garcia, Sergio Gil-Lopez, Daniel Molina, Richard Benjamins, Raja Chatila, and Francisco Herrera.
\newblock Explainable artificial intelligence ({XAI}): Concepts, taxonomies, opportunities and challenges toward responsible {AI}.
\newblock {\em Information Fusion}, 58:82--115, 2020.

\bibitem{blomqvist2009experiments}
Eva Blomqvist, Aldo Gangemi, and Valentina Presutti.
\newblock Experiments on pattern-based ontology design.
\newblock In {\em Proceedings of the Fifth International Conference on Knowledge Capture}, pages 41--48, Redondo Beach, CA, USA, 2009. ACM.

\bibitem{rdfs2014}
Dan Brickley and R.V. Guha.
\newblock {RDF} schema 1.1, 2014.

\bibitem{cardoso2020construction}
Silvio~Domingos Cardoso, Marcos {Da Silveira}, and Cédric Pruski.
\newblock Construction and exploitation of an historical knowledge graph to deal with the evolution of ontologies.
\newblock {\em Knowledge-Based Systems}, 194:105508, 2020.

\bibitem{carriero2021pattern}
Valentina~Anita Carriero, Aldo Gangemi, Maria~Letizia Mancinelli, Andrea~Giovanni Nuzzolese, Valentina Presutti, and Chiara Veninata.
\newblock Pattern-based design applied to cultural heritage knowledge graphs: Arco: The knowledge graph of italian cultural heritage.
\newblock {\em Semantic Web}, 12(2):313--357, 2021.

\bibitem{chen2021augmenting}
Jiaoyan Chen, Ernesto Jim{\'e}nez-Ruiz, Ian Horrocks, Denvar Antonyrajah, Ali Hadian, and Jaehun Lee.
\newblock Augmenting ontology alignment by semantic embedding and distant supervision.
\newblock In {\em The Semantic Web -- ESWC 2021}, pages 392--408, Virtual Event, 2021. Springer.

\bibitem{condeherreros2026llm}
Diego Conde-Herreros, George Hannah, Terry~R. Payne, Jacopo de~Berardinis, Valentina Tamma, David Chaves-Fraga, and Oscar Corcho.
\newblock Llm driven justified ontology alignment.
\newblock In {\em The 3rd Workshop on Evaluation of Language Models in Knowledge Engineering collocated with the 23rd European Semantic Web Conference (ESWC)}, Dubrovnik, Croatia, 2026.

\bibitem{diallo2025ontology}
Abdoulaye Diallo, Claudia d'Amato, and Mouhamadou Thiam.
\newblock Ontology alignment validation using {LLM} and {KG}.
\newblock In {\em The 20th International Workshop on Ontology Matching collocated with the 24th International Semantic Web Conference (ISWC 2025)}, volume 4144, pages 93--100, Nara, Japan, 2025. CEUR-WS.org.

\bibitem{dragisic2016user}
Zlatan Dragisic, Valentina Ivanova, Patrick Lambrix, Daniel Faria, Ernesto Jim{\'e}nez-Ruiz, and Catia Pesquita.
\newblock User validation in ontology alignment.
\newblock In {\em The Semantic Web -- ISWC 2016}, pages 200--217, Kobe, Japan, 2016. Springer.

\bibitem{esnaola2018two}
Iker Esnaola{-}Gonzalez, Jes{\'{u}}s Berm{\'{u}}dez, Izaskun Fern{\'{a}}ndez, and Aitor Arnaiz.
\newblock Two ontology design patterns toward energy efficiency in buildings.
\newblock In {\em Proceedings of the 9th Workshop on Ontology Design and Patterns {(WOP} 2018) co-located with 17th International Semantic Web Conference {(ISWC} 2018)}, volume 2195, pages 14--28, Monterey, CA, USA, 2018. CEUR-WS.org.

\bibitem{euzenat2007semantic}
J{\'e}r{\^o}me Euzenat.
\newblock Semantic precision and recall for ontology alignment evaluation.
\newblock In {\em Proceedings of the 20th international joint conference on Artificial intelligence}, pages 348--353, Hyderabad, India, 2007.

\bibitem{euzenat2007ontology}
J{\'e}r{\^o}me Euzenat and Pavel Shvaiko.
\newblock {\em Ontology Matching (2nd ed.)}.
\newblock Springer, Berlin, Germany, 2013.

\bibitem{faria2014agreementmakerlight}
Daniel Faria, Catia Pesquita, Emanuel Santos, Isabel~F Cruz, and Francisco~M Couto.
\newblock {AgreementMakerLight} 2.0: towards efficient large-scale ontology matching.
\newblock In {\em Proceedings of the ISWC 2014 Posters and Demonstrations Track}, volume 1272, pages 457--460, Riva del Garda, Italy, 2014. CEUR-WS.org.

\bibitem{faria2013agreementmakerlight}
Daniel Faria, Catia Pesquita, Emanuel Santos, Matteo Palmonari, Isabel~F. Cruz, and Francisco~M Couto.
\newblock The {AgreementMakerLight} ontology matching system.
\newblock In {\em On the Move to Meaningful Internet Systems: OTM 2013 Conferences}, pages 527--541, Graz, Austria, 2013. Springer.

\bibitem{brickrec}
Gabe Fierro and Karl Hammar.
\newblock Brick {Schema} and {RealEstateCore} announce a major harmonization effort between two smart building metadata standards, 2022.

\bibitem{gangemi2005ontology}
Aldo Gangemi.
\newblock Ontology design patterns for semantic web content.
\newblock In {\em The Semantic Web -- ISWC 2005}, pages 262--276, Galway, Ireland, 2005. Springer.

\bibitem{gangemi2009ontology}
Aldo Gangemi and Valentina Presutti.
\newblock {\em Ontology Design Patterns}, pages 221--243.
\newblock Springer, Berlin, Germany, 2009.

\bibitem{glimm2014hermit}
Birte Glimm, Ian Horrocks, Boris Motik, Giorgos Stoilos, and Zhe Wang.
\newblock Hermit: An {OWL} 2 reasoner.
\newblock {\em Journal of Automated Reasoning}, 53(3):245--269, 2014.

\bibitem{hammar2019realestatecore}
Karl Hammar, Erik~Oskar Wallin, Per Karlberg, and David H{\"a}lleberg.
\newblock The {RealEstateCore} ontology.
\newblock In {\em The Semantic Web - ISWC 2019}, pages 130--145, Auckland, New Zealand, 2019. Springer.

\bibitem{he2022bertmap}
Yuan He, Jiaoyan Chen, Denvar Antonyrajah, and Ian Horrocks.
\newblock {BERTMap}: A {BERT}-based ontology alignment system.
\newblock In {\em Proceedings of the 36th AAAI Conference on Artificial Intelligence}, volume~36, pages 5684--5691, Virtual Event, 2022. AAAI Press.

\bibitem{he2023exploring}
Yuan He, Jiaoyan Chen, Hang Dong, and Ian Horrocks.
\newblock Exploring large language models for ontology alignment.
\newblock In {\em Proceedings of the ISWC 2023 Posters, Demos and Industry Tracks}, volume 3632, Athens, Greece, 2023. CEUR-WS.org.

\bibitem{he2024language}
Yuan He, Zhangdie Yuan, Jiaoyan Chen, and Ian Horrocks.
\newblock Language models as hierarchy encoders.
\newblock In {\em Advances in Neural Information Processing Systems}, volume~37, pages 14690--14711, Vancouver, Canada, 2024. Curran Associates, Inc.

\bibitem{hertling2023olala}
Sven Hertling and Heiko Paulheim.
\newblock {OLaLa}: Ontology matching with large language models.
\newblock In {\em Proceedings of the 12th Knowledge Capture Conference 2023}, pages 131--139, Pensacola, FL, USA, 2023. ACM.

\bibitem{sweeney2023}
{IMDb}.
\newblock Anyone but you, 2023.

\bibitem{jimenez2011logmap1}
Ernesto Jim{\'e}nez-Ruiz and Bernardo Cuenca~Grau.
\newblock {LogMap}: Logic-based and scalable ontology matching.
\newblock In {\em The Semantic Web -- ISWC 2011}, pages 273--288, Bonn, Germany, 2011. Springer.

\bibitem{jimenez2011logmap2}
Ernesto Jim\'{e}nez-Ruiz, Bernardo~Cuenca Grau, and Yujiao Zhou.
\newblock {LogMap} 2.0: towards logic-based, scalable and interactive ontology matching.
\newblock In {\em Proceedings of the 4th International Workshop on Semantic Web Applications and Tools for the Life Sciences}, pages 45--46, London, UK, 2011. ACM.

\bibitem{doric-mm-2021}
Hedi~M. Karray, Linda Elmhadhbi, and Arkopaul Sarkar.
\newblock Ontologies’ interoperability: Concerns and perspectives, 2021.

\bibitem{klein2001ontology}
Michel Klein and Dieter Fensel.
\newblock Ontology versioning on the semantic web.
\newblock In {\em Proceedings of the First Semantic Web Working Symposium}, pages 75--91, Stanford University, CA, USA, 2001. CEUR-WS.org.

\bibitem{shacl2017}
Holger Knublauch and Dimitris Kontokostas.
\newblock Shapes constraint language ({SHACL}), 2017.

\bibitem{lewis2021retrievalaugmented}
Patrick Lewis, Ethan Perez, Aleksandra Piktus, Fabio Petroni, Vladimir Karpukhin, Naman Goyal, Heinrich K\"{u}ttler, Mike Lewis, Wen-tau Yih, Tim Rockt\"{a}schel, Sebastian Riedel, and Douwe Kiela.
\newblock Retrieval-augmented generation for knowledge-intensive {NLP} tasks.
\newblock In {\em Proceedings of the 34th Annual Conference on Neural Information Processing Systems}, volume~33, pages 9459--9474, Vancouver, Canada, 2020. Curran Associates Inc.

\bibitem{li2021combining}
Guoxuan Li, Songmao Zhang, Jiayi Wei, and Wenqian Ye.
\newblock Combining {FCA-Map} with representation learning for aligning large biomedical ontologies.
\newblock In {\em Proceedings of the 16th International Workshop on Ontology Matching -- ISWC 2021}, volume 3063, pages 207--208, Virtual Event, 2021. CEUR-WS.org.

\bibitem{li2019user}
Huanyu Li, Zlatan Dragisic, Daniel Faria, Valentina Ivanova, Ernesto Jiménez-Ruiz, Patrick Lambrix, and Catia Pesquita.
\newblock User validation in ontology alignment: functional assessment and impact.
\newblock {\em The Knowledge Engineering Review}, 34:e15, 2019.

\bibitem{lushnei2026large}
Sviatoslav Lushnei, Dmytro Shumskyi, Severyn Shykula, Ernesto Jim{\'e}nez-Ruiz, and Artur~d'Avila Garcez.
\newblock Large language models as oracles for ontology alignment.
\newblock In {\em Proceedings of the 19th Conference of the European Chapter of the Association for Computational Linguistics}, pages 2435--2449, Rabat, Morocco, 2026. ACL.

\bibitem{martinez2012logical}
David~Carral Mart\'{\i}nez, Krzysztof Janowicz, and Pascal Hitzler.
\newblock A logical geo-ontology design pattern for quantifying over types.
\newblock In {\em Proceedings of the 20th International Conference on Advances in Geographic Information Systems}, pages 239--248, Redondo Beach, CA, USA, 2012. ACM.

\bibitem{owl2004}
Deborah~L. McGuinness and Frank van Harmelen.
\newblock {OWL} web ontology language overview, 2004.

\bibitem{skos2009}
Alistair Miles and Sean Bechhofer.
\newblock {SKOS} simple knowledge organization system reference, 2009.

\bibitem{s223p}
{National Renewable Energy Laboratory}.
\newblock {ASHRAE Standard 223P}, 2024.

\bibitem{nguyen2025kroma}
Lam Nguyen, Erika Barcelos, Roger French, and Yinghui Wu.
\newblock {KROMA}: Ontology matching with knowledge retrieval and large language models.
\newblock In {\em The Semantic Web -- ISWC 2025}, pages 629--649, Nara, Japan, 2025. Springer.

\bibitem{norouzi2023conversational}
Sanaz~Saki Norouzi, Mohammad~Saeid Mahdavinejad, and Pascal Hitzler.
\newblock Conversational ontology alignment with {ChatGPT}.
\newblock In {\em Proceedings of the 18th International Workshop on Ontology Matching -- ISWC 2023}, volume 3591, pages 61--66, Athens, Greece, 2023. CEUR-WS.org.

\bibitem{noy2001ontology}
Natalya~F. Noy and Deborah~L. McGuinness.
\newblock Ontology development 101: A guide to creating your first ontology.
\newblock Technical Report KSL-01-05/SMI-2001-0880, Stanford University, 2001.

\bibitem{oaei}
{OAEI Community}.
\newblock {Ontology Alignment Evaluation Initiative (OAEI)}, n.d.

\bibitem{chatgpt}
{OpenAI}.
\newblock Introducing {ChatGPT}, 2022.

\bibitem{paulheim2013towards}
Heiko Paulheim, Sven Hertling, and Dominique Ritze.
\newblock Towards evaluating interactive ontology matching tools.
\newblock In {\em The Semantic Web: Semantics and Big Data}, pages 31--45, Montpellier, France, 2013. Springer.

\bibitem{pauwels2016express}
Pieter Pauwels and Walter Terkaj.
\newblock Express to {OWL} for construction industry: Towards a recommendable and usable {ifcOWL} ontology.
\newblock {\em Automation in Construction}, 63:100--133, 2016.

\bibitem{petasis2011ontology}
Georgios Petasis, Vangelis Karkaletsis, Georgios Paliouras, Anastasia Krithara, and Elias Zavitsanos.
\newblock {\em Ontology Population and Enrichment: State of the Art}, pages 134--166.
\newblock Springer, Berlin, Germany, 2011.

\bibitem{plessers2005ontology}
Peter Plessers and Olga De~Troyer.
\newblock Ontology change detection using a version log.
\newblock In {\em The Semantic Web -- ISWC 2005}, pages 578--592, Galway, Ireland, 2005. Springer.

\bibitem{presutti2008content}
Valentina Presutti and Aldo Gangemi.
\newblock Content ontology design patterns as practical building blocks for web ontologies.
\newblock In {\em Conceptual Modeling - ER 2008}, pages 128--141, Barcelona, Spain, 2008. Springer.

\bibitem{qiang2023ontology}
Zhangcheng Qiang.
\newblock Ontology-compliant knowledge graphs.
\newblock In {\em The Semantic Web: ESWC 2023 Satellite Events}, pages 298--309, Hersonissos, Crete, Greece, 2023. Springer.

\bibitem{qiang2025ecosystem}
Zhangcheng Qiang.
\newblock Ontology interoperability: A comprehensive framework for industry-scale applications, 2025.

\bibitem{qiang2023systematic}
Zhangcheng Qiang, Stuart Hands, Kerry Taylor, Subbu Sethuvenkatraman, Daniel Hugo, Pouya~Ghiasnezhad Omran, Madhawa Perera, and Armin Haller.
\newblock A systematic comparison and evaluation of building ontologies for deploying data-driven analytics in smart buildings.
\newblock {\em Energy and Buildings}, 292:113054, 2023.

\bibitem{qiang2024om4ov}
Zhangcheng Qiang, Kerry Taylor, and Weiqing Wang.
\newblock {OM4OV}: Leveraging ontology matching for ontology versioning.
\newblock {\em Transactions on Graph Data and Knowledge}, 4(2):6:1--6:19, 2026.

\bibitem{qiang2024oaei}
Zhangcheng Qiang, Kerry Taylor, Weiqing Wang, and Jing Jiang.
\newblock {OAEI-LLM}: {A} benchmark dataset for understanding large language model hallucinations in ontology matching.
\newblock In {\em Proceedings of the Special Session on Harmonising Generative {AI} and Semantic Web Technologies co-located with the 23rd International Semantic Web Conference}, volume 3953, Baltimore, MD, USA, 2024. CEUR-WS.org.

\bibitem{qiang2023agent}
Zhangcheng Qiang, Weiqing Wang, and Kerry Taylor.
\newblock {Agent-OM}: Leveraging {LLM} agents for ontology matching.
\newblock {\em Proceedings of the {VLDB} Endowment}, 18(3):516--529, 2024.

\bibitem{qiang2025oaei}
Zhangcheng Qiang, Weiqing Wang, and Kerry Taylor.
\newblock {Agent-OM} results for {OAEI} 2025.
\newblock In {\em The 20th International Workshop on Ontology Matching collocated with the 24th International Semantic Web Conference (ISWC 2025)}, volume 4144, pages 202--210, Nara, Japan, 2025. CEUR-WS.org.

\bibitem{sabou2006ontology}
Marta Sabou, Vanessa L{\'{o}}pez, Enrico Motta, and Victoria~S. Uren.
\newblock Ontology selection: Ontology evaluation on the real semantic web.
\newblock In {\em Proceedings of 4th International {EON} Workshop 2006 Evaluation of Ontologies for the Web Co-located with the {WWW2006}}, volume 179, Edinburgh, UK, 2006. CEUR-WS.org.

\bibitem{sassi2016supporting}
Najla Sassi, Wassim Jaziri, and Saad Alharbi.
\newblock Supporting ontology adaptation and versioning based on a graph of relevance.
\newblock {\em Journal of Experimental \& Theoretical Artificial Intelligence}, 28(6):1035--1059, 2016.

\bibitem{sirin2007pellet}
Evren Sirin, Bijan Parsia, Bernardo~Cuenca Grau, Aditya Kalyanpur, and Yarden Katz.
\newblock Pellet: A practical {OWL-DL} reasoner.
\newblock {\em Journal of Web Semantics}, 5(2):51--53, 2007.

\bibitem{sousa2025complex}
Guilherme Sousa, Rinaldo Lima, and Cassia Trojahn.
\newblock Complex ontology matching with large language model embeddings, 2025.

\bibitem{ontologies2009concepts}
Heiner Stuckenschmidt, Christine Parent, and Stefano Spaccapietra.
\newblock {\em Modular Ontologies: Concepts, Theories and Techniques for Knowledge Modularization}.
\newblock Springer, Berlin, Germany, 2009.

\bibitem{taboada2025ontology}
Maria Taboada, Diego Martinez, Mohammed Arideh, and Rosa Mosquera.
\newblock Ontology matching with large language models and prioritized depth-first search.
\newblock {\em Information Fusion}, 123:103254, 2025.

\bibitem{rdf12}
Dominik Tomaszuk and Timothée Haudebourg.
\newblock {RDF} 1.2 schema, 2024.

\bibitem{tsarkov2006fact}
Dmitry Tsarkov and Ian Horrocks.
\newblock Fact++ description logic reasoner: System description.
\newblock In {\em Automated Reasoning, Third International Joint Conference, {IJCAR} 2006 Proceedings}, pages 292--297, Seattle, WA, USA, 2006. Springer.

\bibitem{vardeman2017ontology}
Charles~F. VardemanII, Adila~A. Krisnadhi, Michelle Cheatham, Krzysztof Janowicz, Holly Ferguson, Pascal Hitzler, and Aimee~P.C. Buccellato.
\newblock An ontology design pattern and its use case for modeling material transformation.
\newblock {\em Semantic Web}, 8(5):719--731, 2017.

\bibitem{yang2025language}
Hui Yang, Jiaoyan Chen, Yuan He, Yongsheng Gao, and Ian Horrocks.
\newblock Language models as ontology encoders.
\newblock In {\em The Semantic Web – ISWC 2025: 24th International Semantic Web Conference}, pages 443--461, Nara, Japan, 2025. Springer.

\bibitem{zekri2016tau}
Abir Zekri, Zouhaier Brahmia, Fabio Grandi, and Rafik Bouaziz.
\newblock {$\tau$OWL}: A systematic approach to temporal versioning of semantic web ontologies.
\newblock {\em Journal on Data Semantics}, 5(3):141--163, 2016.

\bibitem{zhang2024large}
Shiyao Zhang, Yuji Dong, Yichuan Zhang, Terry~R. Payne, and Jie Zhang.
\newblock Large language model assisted multi-agent dialogue for ontology alignment.
\newblock In {\em Proceedings of the 2024 International Conference on Autonomous Agents and Multiagent Systems}, pages 2594--2596, Auckland, New Zealand, 2024. IFAAMAS.

\bibitem{zhao2018matching}
Mengyi Zhao, Songmao Zhang, Weizhuo Li, and Guowei Chen.
\newblock Matching biomedical ontologies based on formal concept analysis.
\newblock {\em Journal of Biomedical Semantics}, 9, 2018.

\bibitem{zhou2021towards}
Dongzhuoran Zhou, Baifan Zhou, Jieying Chen, Gong Cheng, Egor Kostylev, and Evgeny Kharlamov.
\newblock Towards ontology reshaping for kg generation with user-in-the-loop: Applied to bosch welding.
\newblock In {\em Proceedings of the 10th International Joint Conference on Knowledge Graphs}, pages 145--150, Virtual Event/Bangkok, Thailand, 2022. ACM.

\bibitem{zhou2022enhancing}
Dongzhuoran Zhou, Baifan Zhou, Zhuoxun Zheng, Egor~V. Kostylev, Gong Cheng, Ernesto Jim{\'e}nez-Ruiz, Ahmet Soylu, and Evgeny Kharlamov.
\newblock Enhancing knowledge graph generation with ontology reshaping -- bosch case.
\newblock In {\em The Semantic Web: ESWC 2022 Satellite Events}, pages 299--302, Hersonissos, Crete, Greece, 2022. Springer.

\bibitem{zhou2022ontology}
Dongzhuoran Zhou, Baifan Zhou, Zhuoxun Zheng, Ahmet Soylu, Gong Cheng, Ernesto Jimenez-Ruiz, Egor~V. Kostylev, and Evgeny Kharlamov.
\newblock Ontology reshaping for knowledge graph construction: Applied on bosch welding case.
\newblock In {\em The Semantic Web -- ISWC 2022}, pages 770--790, Virtual Event, 2022. Springer.

\end{thebibliography}


\begin{thebibliography}{17}
\expandafter\ifx\csname natexlab\endcsname\relax\def\natexlab#1{#1}\fi
\providecommand{\url}[1]{\texttt{#1}}
\providecommand{\href}[2]{#2}
\providecommand{\path}[1]{#1}
\providecommand{\DOIprefix}{doi:}
\providecommand{\ArXivprefix}{arXiv:}
\providecommand{\URLprefix}{URL: }
\providecommand{\Pubmedprefix}{pmid:}
\providecommand{\doi}[1]{\href{http://dx.doi.org/#1}{\path{#1}}}
\providecommand{\Pubmed}[1]{\href{pmid:#1}{\path{#1}}}
\providecommand{\bibinfo}[2]{#2}
\ifx\xfnm\relax \def\xfnm[#1]{\unskip,\space#1}\fi
\bibitem[{Gruber(1993)}]{gruber1993translational}
\bibinfo{author}{T.~R. Gruber},
\newblock \bibinfo{title}{A translational approach to portable ontologies},
\newblock \bibinfo{journal}{Knowledge Acquisition} \bibinfo{volume}{5} (\bibinfo{year}{1993}) \bibinfo{pages}{199--220}. \DOIprefix\doi{10.1006/knac.1993.1008}.
\bibitem[{{IMDb}(2023)}]{sweeney2023}
\bibinfo{author}{{IMDb}}, \bibinfo{title}{Anyone but you}, \bibinfo{year}{2023}. \URLprefix \url{https://www.imdb.com/title/tt26047818/}.
\bibitem[{Gangemi and Presutti(2009)}]{gangemi2009ontology}
\bibinfo{author}{A.~Gangemi}, \bibinfo{author}{V.~Presutti}, \bibinfo{title}{Ontology Design Patterns}, \bibinfo{publisher}{Springer}, \bibinfo{address}{Berlin, Germany}, \bibinfo{year}{2009}, pp. \bibinfo{pages}{221--243}. \DOIprefix\doi{10.1007/978-3-540-92673-3_10}.
\bibitem[{Euzenat and Shvaiko(2013)}]{euzenat2007ontology}
\bibinfo{author}{J.~Euzenat}, \bibinfo{author}{P.~Shvaiko}, \bibinfo{title}{Ontology Matching (2nd ed.)}, \bibinfo{publisher}{Springer}, \bibinfo{address}{Berlin, Germany}, \bibinfo{year}{2013}. \DOIprefix\doi{10.1007/978-3-642-38721-0}.
\bibitem[{Klein and Fensel(2001)}]{klein2001ontology}
\bibinfo{author}{M.~Klein}, \bibinfo{author}{D.~Fensel},
\newblock \bibinfo{title}{Ontology versioning on the semantic web},
\newblock in: \bibinfo{booktitle}{Proceedings of the First Semantic Web Working Symposium}, \bibinfo{publisher}{CEUR-WS.org}, \bibinfo{address}{Stanford University, California, USA}, \bibinfo{year}{2001}, pp. \bibinfo{pages}{75--91}.
\bibitem[{Qiang et~al.(2023)Qiang, Hands, Taylor, Sethuvenkatraman, Hugo, Omran, Perera, and Haller}]{qiang2023systematic}
\bibinfo{author}{Z.~Qiang}, \bibinfo{author}{S.~Hands}, \bibinfo{author}{K.~Taylor}, \bibinfo{author}{S.~Sethuvenkatraman}, \bibinfo{author}{D.~Hugo}, \bibinfo{author}{P.~G. Omran}, \bibinfo{author}{M.~Perera}, \bibinfo{author}{A.~Haller},
\newblock \bibinfo{title}{A systematic comparison and evaluation of building ontologies for deploying data-driven analytics in smart buildings},
\newblock \bibinfo{journal}{Energy and Buildings} \bibinfo{volume}{292} (\bibinfo{year}{2023}) \bibinfo{pages}{113054}. \DOIprefix\doi{10.1016/j.enbuild.2023.113054}.
\bibitem[{Janowicz et~al.(2019)Janowicz, Haller, Cox, {Le Phuoc}, and Lefrançois}]{janowicz2019sosa}
\bibinfo{author}{K.~Janowicz}, \bibinfo{author}{A.~Haller}, \bibinfo{author}{S.~J. Cox}, \bibinfo{author}{D.~{Le Phuoc}}, \bibinfo{author}{M.~Lefrançois},
\newblock \bibinfo{title}{{SOSA}: {A} lightweight ontology for sensors, observations, samples, and actuators},
\newblock \bibinfo{journal}{Journal of Web Semantics} \bibinfo{volume}{56} (\bibinfo{year}{2019}) \bibinfo{pages}{1--10}. \DOIprefix\doi{10.1016/j.websem.2018.06.003}.
\bibitem[{Compton et~al.(2012)Compton, Barnaghi, Bermudez, García-Castro, Corcho, Cox, Graybeal, Hauswirth, Henson, Herzog, Huang, Janowicz, Kelsey, {Le Phuoc}, Lefort, Leggieri, Neuhaus, Nikolov, Page, Passant, Sheth, and Taylor}]{compton2012ssn}
\bibinfo{author}{M.~Compton}, \bibinfo{author}{P.~Barnaghi}, \bibinfo{author}{L.~Bermudez}, \bibinfo{author}{R.~García-Castro}, \bibinfo{author}{O.~Corcho}, \bibinfo{author}{S.~Cox}, \bibinfo{author}{J.~Graybeal}, \bibinfo{author}{M.~Hauswirth}, \bibinfo{author}{C.~Henson}, \bibinfo{author}{A.~Herzog}, \bibinfo{author}{V.~Huang}, \bibinfo{author}{K.~Janowicz}, \bibinfo{author}{W.~D. Kelsey}, \bibinfo{author}{D.~{Le Phuoc}}, \bibinfo{author}{L.~Lefort}, \bibinfo{author}{M.~Leggieri}, \bibinfo{author}{H.~Neuhaus}, \bibinfo{author}{A.~Nikolov}, \bibinfo{author}{K.~Page}, \bibinfo{author}{A.~Passant}, \bibinfo{author}{A.~Sheth}, \bibinfo{author}{K.~Taylor},
\newblock \bibinfo{title}{The {SSN} ontology of the {W3C} semantic sensor network incubator group},
\newblock \bibinfo{journal}{Journal of Web Semantics} \bibinfo{volume}{17} (\bibinfo{year}{2012}) \bibinfo{pages}{25--32}. \DOIprefix\doi{10.1016/j.websem.2012.05.003}.
\bibitem[{Daniele et~al.(2015)Daniele, den Hartog, and Roes}]{daniele2015created}
\bibinfo{author}{L.~Daniele}, \bibinfo{author}{F.~den Hartog}, \bibinfo{author}{J.~Roes},
\newblock \bibinfo{title}{Created in close interaction with the industry: The smart appliances reference {(SAREF)} ontology},
\newblock in: \bibinfo{booktitle}{Formal Ontologies Meet Industry}, \bibinfo{publisher}{Springer}, \bibinfo{address}{Berlin, Germany}, \bibinfo{year}{2015}, pp. \bibinfo{pages}{100--112}. \DOIprefix\doi{10.1007/978-3-319-21545-7_9}.
\bibitem[{Presutti and Gangemi(2016)}]{presutti2016dolce+}
\bibinfo{author}{V.~Presutti}, \bibinfo{author}{A.~Gangemi},
\newblock \bibinfo{title}{Dolce+d{\&}s ultralite and its main ontology design patterns},
\newblock in: \bibinfo{booktitle}{Ontology Engineering with Ontology Design Patterns - Foundations and Applications}, volume~\bibinfo{volume}{25} of \textit{\bibinfo{series}{Studies on the Semantic Web}}, \bibinfo{publisher}{IOS Press}, \bibinfo{address}{Amsterdam, Netherlands}, \bibinfo{year}{2016}, pp. \bibinfo{pages}{81--103}. \DOIprefix\doi{10.3233/978-1-61499-676-7-81}.
\bibitem[{Lebo et~al.(2013)Lebo, Sahoo, McGuinness, Belhajjame, Cheney, Corsar, Garijo, Soiland‑Reyes, Zednik, and Zhao}]{w3c-prov}
\bibinfo{author}{T.~Lebo}, \bibinfo{author}{S.~Sahoo}, \bibinfo{author}{D.~McGuinness}, \bibinfo{author}{K.~Belhajjame}, \bibinfo{author}{J.~Cheney}, \bibinfo{author}{D.~Corsar}, \bibinfo{author}{D.~Garijo}, \bibinfo{author}{S.~Soiland‑Reyes}, \bibinfo{author}{S.~Zednik}, \bibinfo{author}{J.~Zhao}, \bibinfo{title}{{PROV-O}: The {PROV} ontology}, \bibinfo{year}{2013}. \URLprefix \url{https://www.w3.org/TR/prov-o/}.
\bibitem[{{W3C Semantic Sensor Network Incubator Group}(2005)}]{old-ssn}
\bibinfo{author}{{W3C Semantic Sensor Network Incubator Group}}, \bibinfo{title}{Semantic sensor network ontology}, \bibinfo{year}{2005}. \URLprefix \url{https://www.w3.org/2005/Incubator/ssn/ssnx/ssn}.
\bibitem[{Haller et~al.(2017)Haller, Janowicz, Cox, Le~Phuoc, Taylor, and Lefrançois}]{w3c-ssn}
\bibinfo{author}{A.~Haller}, \bibinfo{author}{K.~Janowicz}, \bibinfo{author}{S.~Cox}, \bibinfo{author}{D.~Le~Phuoc}, \bibinfo{author}{K.~Taylor}, \bibinfo{author}{M.~Lefrançois}, \bibinfo{title}{Semantic sensor network ontology}, \bibinfo{year}{2017}. \URLprefix \url{https://www.w3.org/TR/vocab-ssn/}.
\bibitem[{Qiang et~al.(2026)Qiang, Taylor, and Wang}]{qiang2024om4ov}
\bibinfo{author}{Z.~Qiang}, \bibinfo{author}{K.~Taylor}, \bibinfo{author}{W.~Wang},
\newblock \bibinfo{title}{{OM4OV}: Leveraging ontology matching for ontology versioning},
\newblock \bibinfo{journal}{Transactions on Graph Data and Knowledge} \bibinfo{volume}{4} (\bibinfo{year}{2026}) \bibinfo{pages}{6:1--6:19}. \DOIprefix\doi{10.4230/TGDK.4.2.6}.
\bibitem[{Qiang(2023)}]{qiang2023ontology}
\bibinfo{author}{Z.~Qiang},
\newblock \bibinfo{title}{Ontology-compliant knowledge graphs},
\newblock in: \bibinfo{booktitle}{The Semantic Web: ESWC 2023 Satellite Events}, \bibinfo{publisher}{Springer}, \bibinfo{address}{Hersonissos, Crete, Greece}, \bibinfo{year}{2023}, pp. \bibinfo{pages}{298--309}. \DOIprefix\doi{10.1007/978-3-031-43458-7_48}.
\bibitem[{Haller et~al.(2019)Haller, Janowicz, Cox, Lefran{\c{c}}ois, Taylor, Le~Phuoc, Lieberman, Garc{\'\i}a-Castro, Atkinson, and Stadler}]{haller2018modular}
\bibinfo{author}{A.~Haller}, \bibinfo{author}{K.~Janowicz}, \bibinfo{author}{S.~J. Cox}, \bibinfo{author}{M.~Lefran{\c{c}}ois}, \bibinfo{author}{K.~Taylor}, \bibinfo{author}{D.~Le~Phuoc}, \bibinfo{author}{J.~Lieberman}, \bibinfo{author}{R.~Garc{\'\i}a-Castro}, \bibinfo{author}{R.~Atkinson}, \bibinfo{author}{C.~Stadler},
\newblock \bibinfo{title}{The modular {SSN} ontology: {A} joint {W3C} and {OGC} standard specifying the semantics of sensors, observations, sampling, and actuation},
\newblock \bibinfo{journal}{Semantic Web} \bibinfo{volume}{10} (\bibinfo{year}{2019}) \bibinfo{pages}{9--32}. \DOIprefix\doi{10.3233/SW-180320}.
\bibitem[{Taylor et~al.(2019)Taylor, Haller, Lefran{\c{c}}ois, Cox, Janowicz, Garcia-Castro, Le~Phuoc, Lieberman, Atkinson, and Stadler}]{taylor2019semantic}
\bibinfo{author}{K.~Taylor}, \bibinfo{author}{A.~Haller}, \bibinfo{author}{M.~Lefran{\c{c}}ois}, \bibinfo{author}{S.~J. Cox}, \bibinfo{author}{K.~Janowicz}, \bibinfo{author}{R.~Garcia-Castro}, \bibinfo{author}{D.~Le~Phuoc}, \bibinfo{author}{J.~Lieberman}, \bibinfo{author}{R.~Atkinson}, \bibinfo{author}{C.~Stadler},
\newblock \bibinfo{title}{The semantic sensor network ontology, revamped},
\newblock in: \bibinfo{booktitle}{Proceedings of the Journal Track co-located with the 18th International Semantic Web Conference}, volume \bibinfo{volume}{2576}, \bibinfo{publisher}{CEUR-WS.org}, \bibinfo{address}{Auckland, New Zealand}, \bibinfo{year}{2019}.

\end{thebibliography}

\begin{bibunit}

\clearpage
\appendix

\renewcommand{\thesection}{Appendix \Alph{section}}
\renewcommand{\thesubsection}{\Alph{section}.\arabic{subsection}}
\renewcommand{\thesubsubsection}{\Alph{section}.\arabic{subsection}.\arabic{subsubsection}}

\section{Proposed Methodology}

An ontology is designed to achieve semantic interoperability among heterogeneous KGs. However, no ontology is universally applicable. Ontologies are often handcrafted and consistently modified or extended to capture concepts across different domains. This causes an inherent interoperability issue among ontologies, in which concepts may be missing, overlapping, or conflicting. This work introduces a framework for ontology interoperability. We believe that ontology interoperability is not achieved in a single stroke but rather through a process. In each stage of the ontology engineering (OE) life cycle, the use of an appropriate interoperability technique is indispensable.

\begin{enumerate}[wide, noitemsep, topsep=0pt, labelindent=0pt, label=(\roman*)]

\item We review the state-of-the-art semantic web technologies for ontology interoperability. While ontology design patterns (ODPs) focus primarily on the conceptual level, ontology matching and validation (OM\&OV) and ontology validation (OVA) focus primarily on the detailed level of mappings between ontology classes and properties. These techniques also employ different paradigms: ODPs and OVA follow a hub-and-spoke matching schema, whereas OM\&OV uses a pairwise matching schema. The characteristics of these techniques are complementary and offer an opportunity to form a framework for ontology interoperability.

\item We conduct an empirical study of ODPs in the building industry. We find that there is no one-size-fits-all solution in the building domain and that each of the four building ontologies we analysed has advantages and limitations. Therefore, using ODPs provides a means to address interoperability among systems employing different building ontologies. We propose three distinct building ODPs that span macro-level concepts (e.g. building space and systems) to micro-level concepts (e.g. sensors and measurements).

\item We focus on OM\&OV. We first validate that traditional logic-based OM systems are not out-of-date and that several components (e.g. text preprocessing) remain useful in modern OM systems. We then propose a novel agent-powered LLM-based OM system called Agent-OM. Agent-OM overcomes the limitations of the traditional OM system by relying on pre-defined logic and a pre-built knowledge base. By leveraging LLM agents to access background knowledge and external tools, Agent-OM has demonstrated its ability to handle a wide range of OM tasks. Agent-OV extends Agent-OM for OV tasks. We show that OM and OV can share a unified pipeline, but additional modifications are needed to address skewed measurements, update pitfalls, and false mappings. Finally, we show that LLM-based OM systems can be used to construct datasets to evaluate LLM hallucinations on OM tasks.

\item We highlight the importance of OVA. We propose a new paradigm for KGs built from ontologies, the so-called ontology-compliant knowledge graphs (OCKGs). OCKGs not only aim to build ontology-based or ontology-aware KGs but also seek the optimal ontology for given KGs. We present three scenarios in which the ontology needs to fit one KG, two KGs, or multiple KGs with common patterns. OCKGs are a useful tool for data-driven OVA, extending ontology reshaping and ontology learning to support ontology selection and decision-making.

\end{enumerate}

While each semantic web technique focuses on a specific perspective for ontology interoperability, we also identify limitations of each. For example, ODPs are often handcrafted and do not provide detailed mappings between ontology classes and properties; OM\&OV require substantial effort to develop matching systems, and such systems are computationally expensive; and OVA is highly dependent on the output of OM\&OV. Therefore, the appropriate use of each technique at different stages of the OE life cycle is essential to achieve ontology interoperability. From our point of view, we currently position ODPs in the design phase, OM\&OV in the develop phase, and OVA in the deploy phase as the best practice.

\section{Contributions}

\begin{enumerate}[wide, noitemsep, topsep=0pt, labelindent=0pt, label=(\roman*)]

\item We begin with a Socratic questioning ``Is the ontology itself interoperable?'' and highlight the importance of ontology interoperability for data exchange and integration within knowledge graphs (KGs). A running example in the building domain illustrates the effectiveness of using ontology design patterns (ODPs), ontology matching and versioning (OM\&OV), and ontology validation (OVA) across different phases of the ontology engineering (OE) life cycle to build a sustainable ecosystem for ontology interoperability.

\item We provide a comprehensive review of the state-of-the-art semantic web techniques for ontology interoperability, including ontology design patterns (ODPs), ontology matching and versioning (OM\&OV), and ontology validation (OVA). We analyse the advantages and disadvantages of each technique and identify opportunities to integrate them to develop a framework for ontology interoperability~\cite{qiang2025ecosystem}.

\item We systematically analyse four popular ontologies used in the smart building domain and identify the merits and drawbacks of each. We propose building ODPs to capture common and well-agreed concepts and relations in the building domain, including building spaces, equipment and systems, points and measurements~\cite{qiang2023systematic}.

\item We develop a novel software system for OM (Agent-OM~\cite{qiang2023agent}) and an extension for OV (Agent-OV~\cite{qiang2024om4ov}). Agent-OM uses state-of-the-art LLM agents that leverage infrastructure to deliver improved OM performance in a fully-automated and domain-independent manner. Agent-OV extends Agent-OM to explicitly target OV tasks, including new task formulation and evaluation, with a cross-reference mechanism that compiles OM tasks. Agent-OM is used to create an LLM hallucination dataset~\cite{qiang2024oaei} for the LLM leaderboard and LLM fine-tuning for OM. Agent-OM also participates in the Ontology Alignment Evaluation Initiative (OAEI)~\cite{oaei} 2025 campaign and showcases its generative capability in handling OM tasks at various levels of granularity~\cite{qiang2025oaei}.

\item We propose ontology-compliant knowledge graphs (OCKGs) and define three subtasks under OCKGs: ontology compliance within KG, ontology compliance over KGs, and pattern-based compliance. Examples from the building domain demonstrate the process of constraining OCKGs and highlight their efficiency in facilitating ontology validation and decision-making~\cite{qiang2023ontology}.

\item We summarise the contributions made, including several novel solutions for improving ontology design patterns (ODPs), ontology matching and versioning (OM\&OV), and ontology validation (OVA). In addition, we outline ongoing work and future directions for applying semantic web techniques to improve ontology interoperability.

\end{enumerate}

This work explores ontology interoperability at scale. We propose a comprehensive framework that uses ontology design patterns (ODPs), ontology matching and versioning (OM\&OV), and ontology validation (OVA) throughout the ontology engineering (OE) life cycle to improve ontology interoperability. We present several novel approaches for each technique. We expect our work to be an Archimedes' lever that cracks the nuts of ontology interoperability. Our vision for ontology development using our framework runs as follows.

\section{A Guideline for Creating an Interoperable Ontology (Ontology Development 102)}

``Ontology Development 101''~\citep{noy2001ontology} is a well-known guideline for creating an ontology. With the growth of the Semantic Web over the last two decades, ontologies have been used in many downstream applications for data integration and knowledge sharing. This introduces a new requirement to build an interoperable ontology that is easy to transfer and reuse in different ontology-driven applications. Building on top of our comprehensive framework for ontology interoperability, we propose ``Ontology Development 102'', a new guideline for creating an interoperable ontology.

\begin{enumerate}[wide, noitemsep, topsep=0pt, labelindent=0pt, label=(\roman*)]

\item \textbf{Determine domain and scope.} This is an essential step in creating a new ontology. However, we do not expect the competency questions to be complete as we do not create ontology classes, properties, or individuals directly from the competency questions. Instead, competency questions are used to provide a general idea of the domain to facilitate the selection of ODP and ontology candidates.

\item \textbf{Choose an initial ODP.} ODPs can be searched from an ODP corpus, such as \url{https://odpa.github.io/patterns-repository/} (formerly \url{http://ontologydesignpatterns.org}). In practice, a single OPS may not capture all the concepts mentioned in the competency questions, and ODPs can have different levels of abstraction even when they present similar domain concepts. In such cases, it is necessary to integrate multiple ODPs and select ODPs with a suitable level of abstraction based on the competency questions.

\item \textbf{Choose ontology candidates.} Ontology candidates can be searched from the ontology corpus, such as \url{https://ontoportal.org/}. The ontology candidates need to conform to the initial ODP. The ontology candidates that conform to the initial ODPs implicitly conform to the competency questions.

\item \textbf{Find mappings and validate ontology candidates.} This process first uses OM\&OV to align classes, properties, and individuals between ontology candidates. Then, OVA can be used to ensure the mappings are not counter-intuitive and to evaluate the ontology candidates on real KG data.

\item \textbf{Build a new ontology from ontology fragments.} Concepts may be redundant or missing with ontology candidates. In most cases, there is no one ontology that can cover all concepts. We can extract fragments from the ontology candidates and use these ontology fragments to build a new ontology. When merging the ontology fragments, redundant concepts should be removed and missing concepts should be added.

\item \textbf{Revisit and update the initial ODP.} While the new ontology has been validated against the KG data, the initial ODPS should be revisited to update for redundant and missing concepts.

\item \textbf{Iteration process}. Steps (ii)-(vi) can be repeated multiple times until there are no missing or redundant concepts in the updated ODP.

\end{enumerate}

In ``Ontology Development 102'', we suggest creating an ontology not from scratch but by properly referencing existing ODPs and ontologies. This reduces the chance of customised concepts and structural relations being introduced, thereby improving the interoperability of the ontology being created. OM\&OV and OVA provide a solid reference for transferring the ontology from one to another, making it possible for the ontology being created to be interoperable with other downstream applications.

\section{Extended Literature Review}

Ontology interoperability is a fundamental but critical issue in sharing and reusing ontologies across different applications. While various techniques have been applied in this area, this paper provides a review of three dominant state-of-the-art semantic web techniques for ontology interoperability: ontology design patterns (ODPs), ontology matching and versioning (OM\&OV), and ontology validation (OVA). We analyse the advantages and disadvantages of each technique and identify opportunities to integrate them to build a comprehensive framework for ontology interoperability.

There is no doubt that using an ontology for knowledge graphs (KGs) can enhance KG interoperability through shared vocabularies and definitions, but the ontology itself is not interoperable. For example, Brick~\citep{balaji2016brick} and RECore~\citep{hammar2019realestatecore} are two popular ontologies in the building domain, but their coverage is different and they do not share the same concepts. Brick offers more detailed concepts for building sensors and actuators, while RECore is better at presenting spatial information about the building. The two ontologies also use different terminologies to describe identical concepts. For example, the building's physical location is referred to as ``Site'' in Brick and ``Land'' in RECore. A proposal~\citep{brickrec} to harmonise these two ontologies began in 2022, but to date, many concepts have remained unaligned. Moreover, this collaboration not only aims to align two building ontologies but also includes many other emerging alignments, such as the Brick alignment to two industry standards, Industry Foundation Classes (IFC)~\citep{pauwels2016express} and ASHRAE Standard 223P~\citep{s223p}.

Ontology interoperability is an essential requirement for ontology engineering. It is not possible for two applications that use different ontologies to communicate without aligning their concepts. Several research areas, such as ontology modularisation~\citep{ontologies2009concepts} and ontology selection~\citep{sabou2006ontology}, are highly dependent on the level of ontology interoperability. When a replacement cannot be found for an ontology entity, the extracted ontology module may treat it as missing. This will lead to an entirely different choice of the ontology.

We provide a comprehensive review of state-of-the-art semantic web techniques for ontology interoperability. These include ontology design patterns (ODPs), ontology matching and versioning (OM\&OV), and ontology validation (OVA). We review the current literature, identify knowledge gaps, and outline future directions. By leveraging their strengths while mitigating their weaknesses, we identify opportunities to build a comprehensive framework across these three techniques to improve ontology interoperability.

\subsection{Preliminaries}
\label{sec: preliminaries}

\subsubsection{Ontology}

An ontology contains classes, properties, and individuals. In this work, the word \textit{entity} (or equivalently \textit{term}) is a general expression for ontology classes, object properties, datatype properties, or individuals (without specifying which). We use \textit{entity uri} to mean a fully expanded entity identifier with respect to its prefix. We use \textit{entity name} to mean an entity identifier trimmed of its prefix.

\begin{figure}[htbp]
\centering
\includegraphics[width=1\columnwidth]{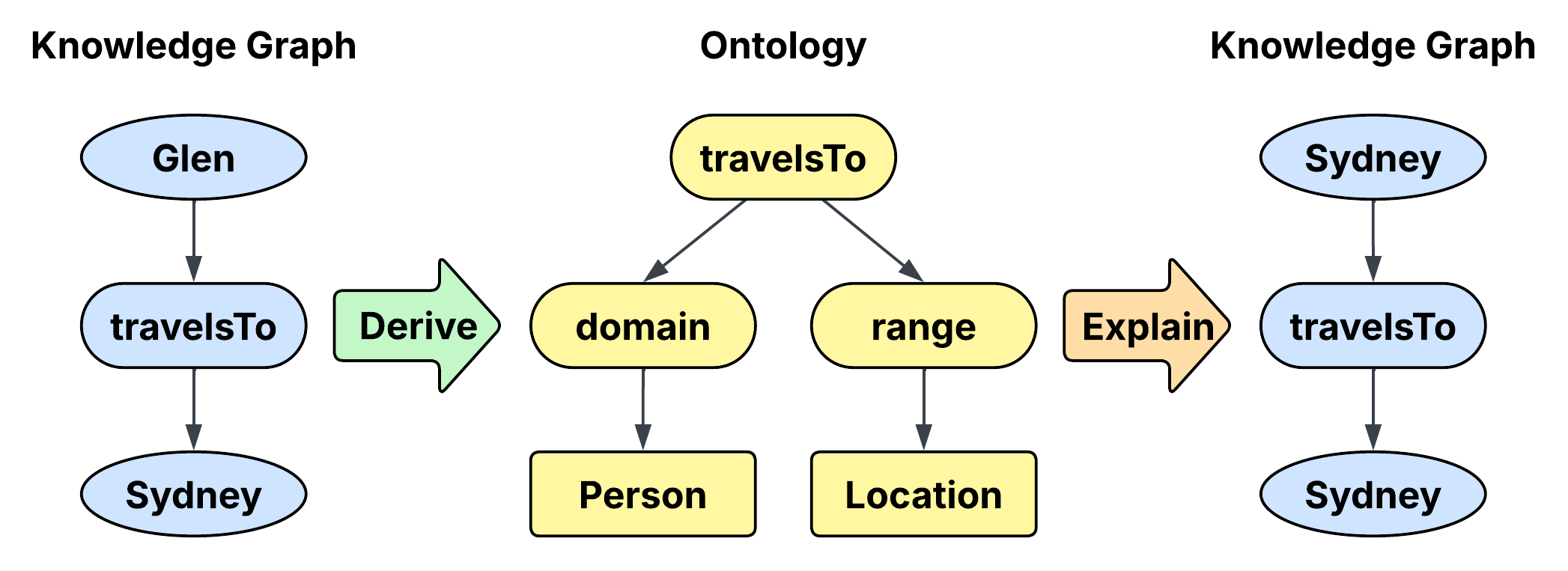}
\caption{An example of using an ontology to explain a knowledge graph.}
\label{fig: ontology-kg}
\end{figure}

The ontology is the cornerstone of KGs. KGs built from an ontology are much easier to interpret and understand. Figure~\ref{fig: ontology-kg} shows an example of using an ontology for KGs. The KG describes an event where Glen Powell and Sydney Sweeney travel to the city of Sydney to film a romantic comedy movie ``Anyone But You''~\citep{sweeney2023}. The triples (Glen, travelsTo, Sydney) and (Sydney, travelsTo, Sydney) are triples of the KG, while the triples (travelsTo, domain, Person) and (travelsTo, range, Place) are triples of the ontology. The ontology triples derived from the KG triple (Glen, travelsTo, Sydney) can be used to explain the perplexing KG triple (Sydney, travelsTo, Sydney). By using definitions provided by the ontology, it is easier to infer that the first subject entity ``Sydney'' refers to the person Sydney Sweeney, while the second object entity ``Sydney'' refers to the city of Sydney.

\begin{figure}[!t]
\centering
\includegraphics[width=1\columnwidth]{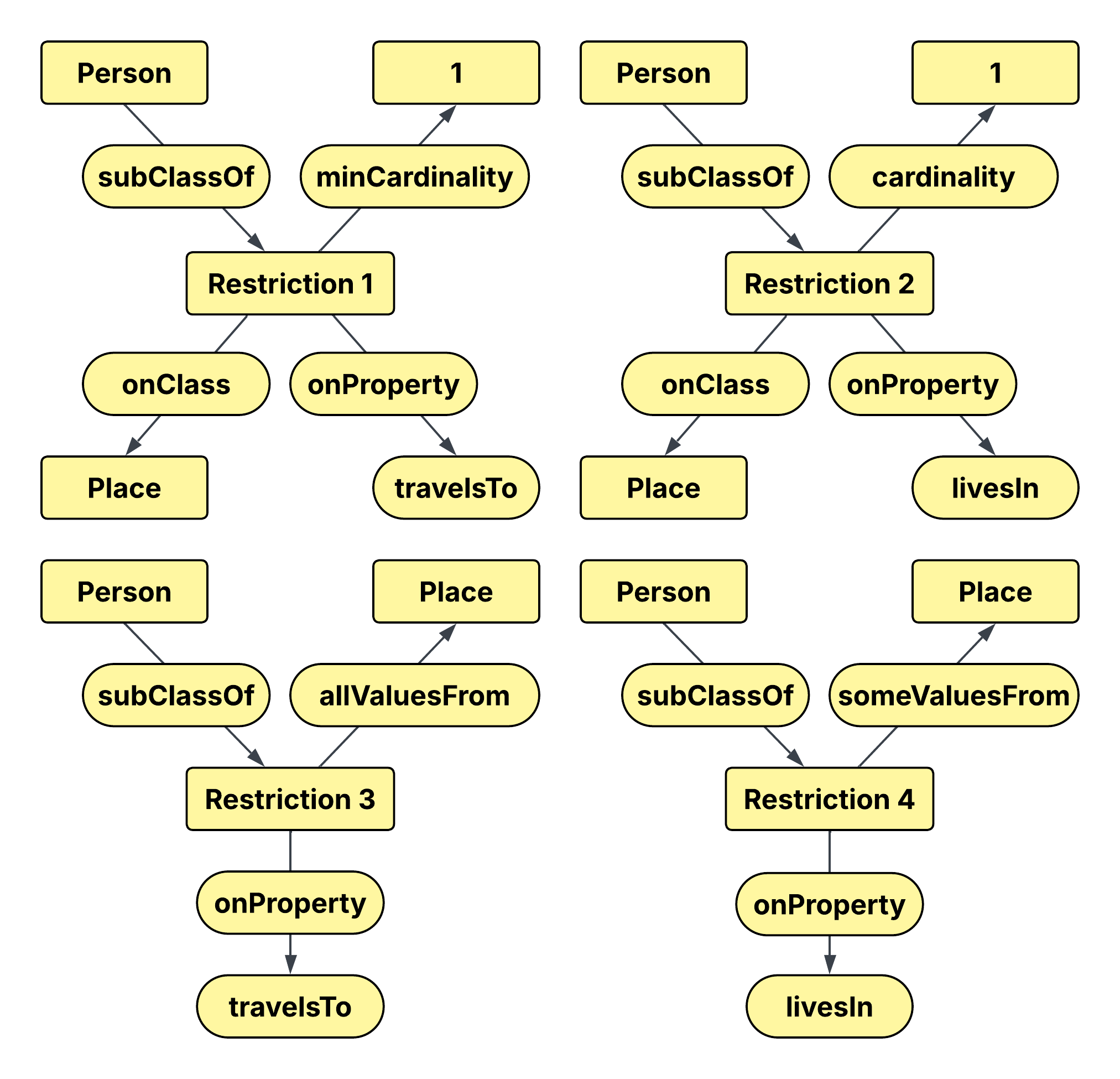}
\caption{Examples of using logical expressions within the ontology.}
\label{fig: ontology-restriction}
\end{figure}

The ontology also provides extensive logical expressions for KGs to represent implicit knowledge from which explicit consequences may be deductively inferred, and to represent constraints on explicit knowledge that may be used to detect violations that are considered to be errors in the KGs. Figure~\ref{fig: ontology-restriction} shows several examples of using logical expressions within the ontology. A person can travel to many places (Restriction 1), but can only live in one place (Restriction 2). If a person travels, they travel to a place (Restriction 3). Every person lives in a place (Restriction 4). These pieces of information cannot be represented in KGs, but can be modelled as machine-readable restrictions within the ontology.

Developing an ontology is not easy and requires extensive domain knowledge, but once it is developed, the ontology can be used across multiple KGs. Both the ontology and the KGs form a set of triples (subject, predicate, object). Still, ontology provides term definitions for entity types (e.g. class, property, individual) as well as more advanced logical expressions (e.g. domain, range, and restriction). These terms can be modelled and formatted in the Web Ontology Language (OWL)~\citep{owl2004} and other languages, such as Resource Description Framework Schema (RDFS)~\citep{rdfs2014} and Simple Knowledge Organisation System (SKOS)~\citep{skos2009}, both of which are strongly tied to OWL.

In general, the ontology serves as the conceptual model, whereas KGs store data instances. There is one exception: a data instance can be defined as an individual within the ontology, but its existence also means that the data instance (i.e. individual) is a well-agreed term that is not always changeable. For example, the term ``Sydney'' can be defined as an individual within the ontology. KGs do not always require an ontology. For some domains, ontology-free KGs are a more rational choice compared to ontology-based KGs. For example, building an ontology for social media is not recommended due to the rapid evolution of terms and the lack of universal agreement. In this context, ``tags'' are commonly used as metadata to describe the context of utterances.

\subsubsection{Ontology Interoperability}

Ontologies have been used for decades to improve the interoperability of KGs and relational databases. However, a criticised view is that the ontology itself is not interoperable. An ontology can be built from scratch or reused from existing ontologies, but there is no guarantee that an ontology always contains the same concepts as other ontologies. Overlaps and missing concepts may exist, making it difficult to share and reuse an ontology in different applications. There are various types of interoperability issues in the ontology~\citep{doric-mm-2021}.

\begin{enumerate}[wide, noitemsep, topsep=0pt, labelindent=0pt, label=(\roman*)]
\item Formatting interoperability. The ontology can be expressed as OWL, RDF, or XML. There is a known issue with transferring ontologies to different languages due to differences in their levels of expressiveness.
\item Terminological interoperability. The ontology can use different terminologies to describe the same concepts. For example, ``AHU'' and ``Air Handling Unit''.
\item Logical interoperability. The same concepts can have different meanings in different ontologies. For example, if the class ``Meter'' is the subclass of ``Point'', it means a data point; if the class ``Meter'' is the subclass of ``Device'', it means a physical device.
\item Semantic interoperability. The same concepts can be modelled from different perspectives. For example, the class ``HVAC Zone'' can be logically considered as a functional unit or physically as a concrete spatial element.
\item Contextual interoperability. The same concepts can have different meanings in different domains. For example, the class ``System'' refers to a software application in the building domain, but means a set of organs in human anatomy.
\end{enumerate}

\subsection{Ontology Design Patterns}
\label{sec: odps}

\subsubsection{Overview}

Ontology design patterns (ODPs) are built on the concept of modularisation and reusability. ODPs are the smallest and most reusable building blocks for constructing large ontologies~\citep{gangemi2005ontology,gangemi2009ontology,blomqvist2009experiments}. ODPs are extracted or re-engineered from existing ontologies. For example, the Actuation-Actuator-Effect (AAE) ODP is extracted from SAN~\citep{san}, while the Execution-Executor-Procedure (EEP) ODP is re-engineered from SEAS~\citep{seas}, SOSA~\citep{sosa}, SSN~\citep{ssn}, and DUL~\citep{dul}. Ontologies that follow the same ODP are interoperable. One can easily interpret identical ontology classes, properties, and individuals on the basis of shared patterns.

\begin{figure}[htbp]
\centering
\includegraphics[width=1\columnwidth]{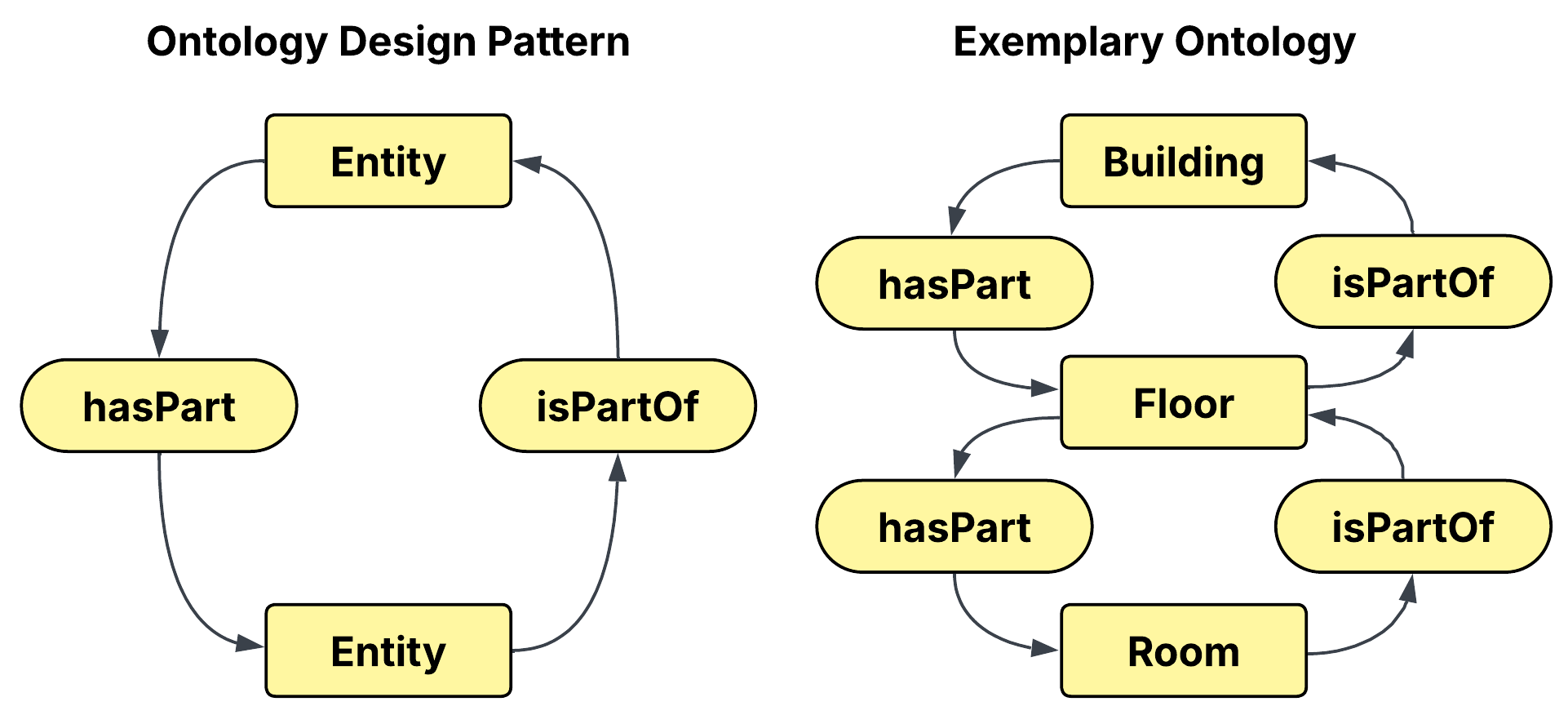}
\caption{Ontology design pattern vs exemplary ontology.}
\label{fig: ODP-Exemplary}
\end{figure}

ODPs are not fully implemented as a real ontology and provide only high-level information on the design of the ontology. The fully implemented ontology is commonly referred to as an exemplary ontology, which is often confused with ODPs. Figure~\ref{fig: ODP-Exemplary} shows an example of using the ``part of'' pattern in the ODP and the exemplary ontology for the building domain. The ODP contains only one ``hasPart'' and one ``isPartOf'', but the exemplary ontology repeats these terms multiple times to model domain-specific part-whole relationships among buildings, floors, and rooms.

\subsubsection{Ontology Design Patterns in the Wild}

ODPs have different types, including content ODPs, structural ODPs for logical and architectural issues, correspondence ODPs for reengineering and alignment, and other types such as reasoning, presentation, and lexico-syntactic ODPs. Content ODPs are the most widely used ODPs, defining patterns of classes, properties, and relations for domain ontologies~\citep{presutti2008content}.

One of the most well-known ODP repositories is available at \url{https://odpa.github.io/patterns-repository/} (formerly \url{http://ontologydesignpatterns.org}), with documentation on domain usage and competency questions. ODPs have been used in many domains, such as bioinformatics~\citep{aranguren2008ontology}, geography~\citep{martinez2012logical}, cultural heritage~\citep{carriero2021pattern}, material science~\citep{vardeman2017ontology}, and energy efficiency in buildings~\citep{esnaola2018two}.

\subsubsection{Knowledge Gaps}

\begin{enumerate}[wide, noitemsep, topsep=0pt, labelindent=0pt, label=(\roman*)]

\item Content ODPs dominate the current ODPs. We do not see many of the other types of ODPs available, but they can be useful in some cases.

\item Limited research is available on using machine learning (ML) or large language models (LLMs) to facilitate this process. The major problem is determining the level of abstraction for each ODP. Even with a number of samples provided, it seems challenging to apply a few-shot approach to train a model for this purpose.

\item Current ODPs are handcrafted by humans and validated on a case-by-case basis. This makes it challenging to apply ODPs in complex and emerging domains. For example, the building domain encompasses a mixture of building spaces, equipment and systems, points and measurements.

\end{enumerate}

\subsection{Ontology Matching and Versioning}
\label{sec: omov}

\subsubsection{Overview}

Ontology matching (OM) aims to find similar concepts in different ontologies~\citep{euzenat2007ontology}. OM is also known as ontology alignment (OA), but OA is more commonly used to describe the OM process. Ontology versioning (OV) aims to capture the differences between different versions of an ontology~\citep{klein2001ontology}. It is easy to see that OV is a specific form of OM. In some contexts, ontology evolution is used as a synonym for OV, but OV is preferred because ontologies are not evolved naturally, but rather through human design. Ontologies are interoperable when their classes, properties, and individuals are explicitly aligned with OM\&OV. One can easily transfer terms from one ontology to another using a reference alignment.

Given a pair of ontologies, the OM\&OV task can be formed by finding an alignment that contains a set of mappings~\citep{euzenat2007semantic}. OM\&OV tasks target equivalence commonly, as other relations (e.g. subsumption or disjointness) can be derived from equivalent mappings. For example, if A is equivalent to B and B is a subclass of C, then A should also be a subclass of C. Similarly, the OM\&OV task can be more complex and extended to one-to-many and many-to-many mappings, but the most common OM\&OV tasks target one-to-one mappings.

\subsubsection{Ontology Matching}

\begin{figure}[htbp]
\centering
\includegraphics[width=1\columnwidth]{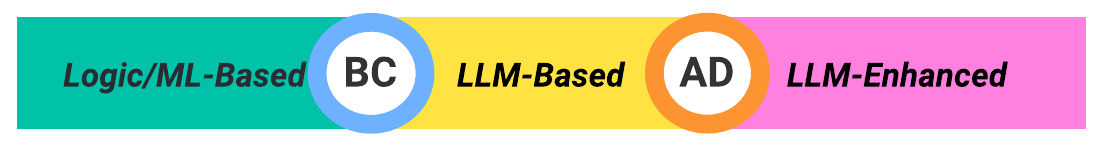}
\caption{Roadmap of ontology matching (OM).}
\label{fig: roadmap-om}
\end{figure}

Figure~\ref{fig: roadmap-om} illustrates the roadmap of OM. Before ChatGPT (BC), a number of OM systems were developed using traditional logic-based expert systems and ML-based prediction systems. Following the introduction of ChatGPT in 2022~\citep{chatgpt}, there was a brief period of using a purely LLM-based approach to perform OM tasks. Due to the current limitations of LLMs, the period after destress (AD) aims to use LLMs with enhancements.

\begin{enumerate}[wide, noitemsep, topsep=0pt, labelindent=0pt, label=(\roman*)]

\item Logic-based systems are highly dependent on domain expertise, and mapping is mainly based on the interpretation of logic expressions within the ontology. Typical logic-based OM systems include LogMap~\citep{jimenez2011logmap1}~\citep{jimenez2011logmap2}, AgreementMakerLight (AML)~\citep{faria2013agreementmakerlight}~\citep{faria2014agreementmakerlight}, and FCA-Map~\citep{zhao2018matching}~\citep{li2021combining}. Such systems generally compile the results quickly with acceptable performance for simple OM tasks. However, these systems face a bottleneck that prevents them from performing complex OM tasks. It is not known how to capture a comprehensive list of logic expressions in various domains and applications.

\item ML-based systems are derived from the popular concept of machine learning in the field of artificial intelligence (AI). LogMap-ML~\citep{chen2021augmenting} and BERTMap~\citep{he2022bertmap} are successful examples. Such systems typically include a learning process that requires a certain amount of training data for the model to learn patterns and make predictions. As with ML applications in other domains, ML-based systems face the challenge of sampling high-quality training data from the full dataset. The explainability of model choice is a long-standing issue within the scope of explainable AI (XAI)~\citep{arrieta2020explainable}.

\item LLM-based systems refer to systems that use LLMs directly for OM. In the early stages of using LLMs for OM, a general approach is to feed ontologies into the LLM entirely with prompt instructions. Such examples can be found in~\citep{he2023exploring} and~\citep{norouzi2023conversational}. The scalability of this approach is problematic. For large-scale ontologies, it may not be feasible to feed the entire ontology into LLMs due to input token limits. In addition, these systems often exhibit low precision and recall due to the inherent hallucinations associated with LLMs.

\item LLM-enhanced systems hold the view that current LLMs are not perfect. It is required to re-engineer the LLM-based infrastructure for OM tasks. This field is a new research area with two main streams. One is using retrieval-augmented generation (RAG)~\citep{lewis2021retrievalaugmented} to inject external resources and extend the knowledge base of LLMs. This often results in the introduction of novel LLM-based architectures for OM tasks, such as~\citep{hertling2023olala,giglou2024llms4om,zhang2024large,taboada2025ontology,nguyen2025kroma}. The other stream aims to feed external knowledge into LLMs by training or fine-tuning, as in~\citep {he2024language,sousa2025complex,yang2025language}.

\end{enumerate}

Table~\ref{tab: om-system-type} summarises the pros and cons of different types of OM systems. While LLM-enhanced architectures are the state-of-the-art solution for OM systems, they also suffer from high cost and non-determinism, which can be mitigated by integrating traditional system designs.

\begin{table*}
\centering
\renewcommand\arraystretch{1.2}
\tabcolsep=0.15cm
\caption{Pros and cons of different types of ontology matching systems.}
\label{tab: om-system-type}
\begin{adjustbox}{width=\textwidth,center}
\begin{tabular}{|m{2.5cm}|m{7.5cm}|m{7.5cm}|} \hline
\textbf{System Type} & \textbf{Pros} & \textbf{Cons} \\ \hline
Logic-Based
&\begin{itemize}[left=0pt]
\item Acceptable performance for simple tasks
\item Fast and requires low computational power
\end{itemize}
&\begin{itemize}[left=0pt]
\item Requires extensive domain expertise
\item Meet the bottleneck for complex OM tasks
\end{itemize}
\\ \hline
ML-Based
&\begin{itemize}[left=0pt]
\item High accuracy in trained models
\item Models can be generalised for different domains
\end{itemize}
&\begin{itemize}[left=0pt]
\item Lack of high-quality training data
\item Lack of explainability
\end{itemize}
\\ \hline
LLM-Based
&\begin{itemize}[left=0pt]
\item Ease of use with intuitive prompt engineering
\item Provides mapping explanations
\end{itemize}
&\begin{itemize}[left=0pt]
\item Not scalable for large-scale ontologies
\item Low accuracy caused by LLM hallucinations
\end{itemize}
\\ \hline
LLM-Enhanced
&\begin{itemize}[left=0pt]
\item High accuracy in complex tasks
\item Provides precise mapping explanations
\end{itemize}
&\begin{itemize}[left=0pt]
\item Extensive burden for system re-engineering
\item Dynamic outputs due to the LLM non-determinism
\end{itemize}
\\ \hline
\end{tabular}
\end{adjustbox}
\end{table*}

\subsubsection{Ontology Versioning}

\begin{figure}[htbp]
\centering
\includegraphics[width=1\columnwidth]{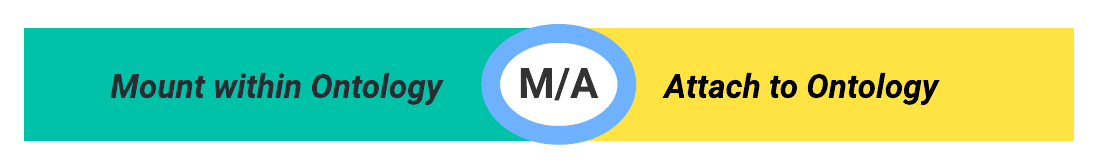}
\caption{Roadmap of ontology versioning (OV).}
\label{fig: roadmap-ov}
\end{figure}

Figure~\ref{fig: roadmap-ov} illustrates that there are two dominant approaches to OV tasks. One approach is to document versioning information within the ontology and expose it as additional triples. The other approach is to document the versioning information outside the ontology, as a complement to the ontology.

\begin{enumerate}[wide, noitemsep, topsep=0pt, labelindent=0pt, label=(\roman*)]

\item Versioning information within the ontology requires an extension of the current vocabulary. OWL has adopted several annotation properties related to versioning information. These include version tags (versionInfo and priorVersion), compatible tags (backwardCompatibleWith and incompatibleWith), and deprecated tags (DeprecatedClass and DeprecatedProperty). On the other hand, $\tau$OWL~\citep{zekri2016tau} and RDF 1.2~\citep{rdf12} have extended a fourth dimension within the quadruples to capture versioning information. However, this approach can consistently increase the size of the ontology, making it unsuitable for large and rapidly evolving ontologies.

\item Attaching versioning information to the ontology can be done in many ways. Versioning information can be captured as a log~\citep{plessers2005ontology}, a standalone KG~\citep{cardoso2020construction}, or a network of KGs~\citep{sassi2016supporting}. While this approach is more flexible in its support for formats and does not increase the size of the ontology, it requires additional control and stores versioning files.

\end{enumerate}

\subsubsection{Knowledge Gaps}

\begin{enumerate}[wide, noitemsep, topsep=0pt, labelindent=0pt, label=(\roman*)]

\item In the era of LLMs, LLMs offer extensive capabilities to perform OM tasks. It is questionable whether the logic and ML used in classical OM systems remain useful in modern LLM-based systems.

\item LLMs are useful tools for OM, but OM systems are complex and lack centralised control, which cannot meet the requirements for large-scale automation. LLM agents could be an interesting area to explore, and LLM hallucinations need to be properly understood and mitigated.

\item OV is similar to OM, but it is unclear whether the pipeline for OM tasks can be reused for OV tasks. There is no unified pipeline that combines OM and OV.

\end{enumerate}

\subsection{Ontology Validation}
\label{sec: ova}

\subsubsection{Overview}

Ontology validation (OVA) aims to check the compliance of the ontology. OVA is sometimes confused with ontology verification (OVF), but the two have different goals. The former refers to ``Do we build the ontology correctly?'', while the latter refers to ``Do we build the correct ontology?''. OVF ensures that the ontology is intrinsically correct, while OVA ensures that the ontology is extrinsically meaningful.

\subsubsection{Ontology Matching Validation}

OM outputs cannot be directly used for downstream tasks. It is essential to include a validation step to ensure that the matching outputs align with real-world facts. This process is also known as ``matching repair'', which aims to identify counter-intuitive mappings that are mistakenly detected by OM systems. OM validation can be classified into two categories.

\begin{enumerate}[wide, noitemsep, topsep=0pt, labelindent=0pt, label=(\roman*)]
\item Logic-based validation. The traditional approach is to use reasoners (e.g. FaCT++~\citep{tsarkov2006fact}, Pellet~\citep{sirin2007pellet}, and HermiT~\citep{glimm2014hermit}) to perform a coherence check on the mappings. OM outputs are formed as assertions and combined with the original ontology. A mapping is incorrect if the reasoner finds that the internal rules have been violated. Recently, this validation has been extended to use LLM prompts to verify mapping correctness~\citep {diallo2025ontology,lushnei2026large,condeherreros2026llm}.
\item Human-in-the-loop validation. This approach uses human experts to validate the OM outputs. Common methods include surveys, interviews, and workshops. Several studies~\citep{paulheim2013towards,dragisic2016user,li2019user} have shown that the effectiveness of human-in-the-loop validation is highly dependent on the user groups selected, the user interface used, and the instructions provided.
\end{enumerate}

\subsubsection{Data-Driven Ontology Validation}

One common type of OVA is to apply ontologies to a specific task or application and validate them using data instances. This process is known as data-driven ontology validation (DDOV). Ontologies that conform to the same data are interoperable. One can easily determine mappings between ontologies if entities are assigned to the same instance. KGs are the data most commonly used for DDOV. Depending on different assumptions and baselines, DDOV can be categorised into two types: ontology-based KGs and ontology-aware KGs. Ontology-based KGs assume the ontology is complete, and KGs need to be modified if they do not conform to the ontology. On the other hand, ontology-aware KGs treat KGs as baselines, and the ontology needs to be modified if it does not conform to them. In the literature, modifying the ontology of an ontology-aware KG is also referred to as ontology enrichment~\citep{petasis2011ontology} and ontology reshaping~\citep{zhou2021towards,zhou2022enhancing,zhou2022ontology}.

Ontology-based KGs validate KGs based on the given ontology. For example, when the domain of ``Feature of Interest'' is defined as observation or actuation in the ontology, linking this term to a sensor or actuator violates this rule. Shapes Constraint Language (SHACL)~\citep{shacl2017} is the W3C recommendation to validate KGs that comply with the ontology. SHACL provides more comprehensive and extensible validation than the built-in logical expressions in the ontology. SHACL allows KGs to conform to a specific ``shape'' defined by the end user. This ``shape'' can be a simple triple, but it can also be a complex shape graph.

Ontology-aware KGs validate the ontology against the given KGs, identifying missing or unused concepts. Ontology enrichment is the process of adding missing concepts to the ontology. For example, if the KG contains ``Quality'' under ``Feature of Interest'', then the ontology needs to add this new concept accordingly. Ontology reshaping is the process of removing unused concepts from the ontology. For example, if the KG is only related to sensor observation, then ``Trigger'' used for an actuator can be removed from the ontology.

\subsubsection{Knowledge Gaps}

\begin{enumerate}[wide, noitemsep, topsep=0pt, labelindent=0pt, label=(\roman*)]

\item OM validation is labour-intensive. It is not easy to find domain experts and they may also have biases. Using an LLM as an alternative is feasible but introduces the risk of LLM hallucinations, which can reduce the quality of OM validation.

\item In reality, ontologies are inherently incomplete. Although ontology-based KGs have been applied to many domains, they are not always applicable, especially in emerging domains where concepts are evolving rapidly. While ontology enrichment adds new concepts and ontology reshaping deletes unused concepts, there is a need to support adding, deleting, and updating functions concurrently. Moreover, ontology-aware KGs assume that the selected ontology is the optimal choice for the given KGs. Limited attention has been paid to finding an optimal ontology from a number of candidates or pattern-based ontology fragments.

\item There are no standard evaluation metrics and datasets for OVA. This limitation makes it difficult to choose an ontology. For example, the decision can vary when one concept is missing from one ontology and misinterpreted in another.

\end{enumerate}

\subsection{Discussion}
\label{sec: discussion}

Table~\ref{tab: compare-technique} compares three state-of-the-art semantic web techniques for ontology interoperability. We can see that ODPs, OM\&OV, and DDOV differ from several perspectives, including prerequisite, concept coverage, alignment approach and schema, and medium used.

\begin{enumerate}[wide, noitemsep, topsep=0pt, labelindent=0pt, label=(\roman*)]

\item ODPs provide high-level guidelines for building ontologies. ODPs achieve ontology interoperability by aligning the conceptual model. This approach often extracts common patterns across different ontologies in a hub and ignores their differences. ODPs can be reused across different ontologies.

\item OM\&OV aim to find the corresponding mappings between two ontologies. OM\&OV achieve ontology interoperability by aligning ontology entities. While OM only reports aligned entities between ontologies (i.e. commonalities), OV also reports unaligned entities (i.e. differences).

\item OVA validates the ontologies to align with human values and data instances. OM validation serves as a prerequisite for DDOV, and DDOV achieves ontology interoperability by aligning the data instances. This approach typically applies a KG hub across multiple ontology spokes. Both OM validation and DDOV process need reference alignments to identify commonalities between ontologies and thereby highlight their differences.

\end{enumerate}

These three techniques can work independently for ontology interoperability, but they also have some inter-dependencies. Therefore, there are opportunities to develop a framework that leverages their strengths while mitigating their weaknesses.

\begin{enumerate}[wide, noitemsep, topsep=0pt, labelindent=0pt, label=(\roman*)]

\item ODPs provide only high-level guidelines for conceptual modelling and lack detailed mappings between classes, properties, and individuals. ODPs often need to cooperate with OM\&OV and OVA in downstream tasks.

\item  OM\&OV provide detailed mappings for classes, properties, and individuals, but with limited support from the modelling perspective. ODPs can simplify the OM\&OV process. Ontologies that use the same ODP are much easier to align, as they share the same conceptual model. The output of OM\&OV can serve as a reference for OVA.

\item OVA is an evaluation process and does not provide any mappings by nature. OVA reuses the commonalities generated by OM\&OV and thereby identifies missing or overlapping modelling issues. The output of OVA can be used to optimise existing ODPs in the next cycle. For example, removing overlap and adding missing concepts and relations.

\end{enumerate}

\begin{table*}
\centering
\renewcommand\arraystretch{1.2}
\tabcolsep=0.15cm
\caption{Comparison of three state-of-the-art semantic web techniques for ontology interoperability: ontology design patterns (ODPs), ontology matching and versioning (OM\&OV), and ontology validation (OVA).}
\label{tab: compare-technique}
\begin{adjustbox}{width=\textwidth,center}
\begin{tabular}{|c|c|c|c|c|c|} \hline
\textbf{Technique} & \textbf{Prerequisite}   & \textbf{Concept Coverage}    & \textbf{Alignment Approach}   & \textbf{Alignment Schema} & \textbf{Medium Used}    \\ \hline
ODPs               & Ontologies Only         & Commonalities Only           & Align Conceptual Model        & Hub-and-Spoke             & Common Patterns         \\ \hline
OM\&OV             & Ontologies Only         & Commonalities + Differences  & Align Ontology Entities       & Pairwise                  & N/A                     \\ \hline
OVA               & Ontologies + References & Differences Only              & Align human values/instances  & Hub-and-Spoke             & Knowledge Graphs        \\ \hline
\end{tabular}
\end{adjustbox}
\end{table*}

While the introduction of ontologies aims to enable interoperability among KGs, with numerous ontologies developed for different domains and purposes, ontology interoperability is becoming a bottleneck for sharing and reusing ontologies in real-world applications. In this paper, we review three dominant semantic web techniques for ontology interoperability. We observe that each targets a specific perspective on ontology interoperability and differs in strengths and weaknesses. The inter-dependencies between ODPs, OM\&OV, and OVA offer opportunities to develop a comprehensive framework for ontology interoperability.

\paragraph*{Supplemental Material Statement}

The source code, data, and/or other artifacts used in this study are available at \url{https://github.com/qzc438-research/ontology-interoperability}.

\putbib[qiang-bibliography]

\end{bibunit}

\end{document}